\documentclass[]{aa} 
\usepackage{graphicx}
\usepackage{txfonts}
\begin{document}
\title{Permissible transitions of the variability classes of GRS 1915+105}

   \author{Partha Sarathi Pal \inst{1}, Sandip K. Chakrabarti \inst{2} \inst{1} 
           \and Anuj Nandi \inst{3}
         }

   \institute{Indian Centre For Space Physics, 
              43 Chalantika, Garia Station Road, Kolkata - 700084, India\\
              \email{partha@csp.res.in, parthasarathi.pal@gmail.com}
         \and
             S. N. Bose National Centre for Basic Sciences,
             JD Block, Salt lake, Kolkata - 700094, India.\\
             \email{chakraba@bose.res.in}
          \and
             Indian Space Research Organization, Bangalore, India\\
              \email{anuj@isac.gov.in}
             }

   \date{Received ; accepted }

 
  \abstract
{The Galactic microquasar GRS 1915+105 exhibits at least sixteen types of variability classes. Transitions 
from one class to another could take place in a matter of hours. In some of the classes, the spectral state
transitions (burst-off to burst-on and vice versa) were found to take place 
in a matter of few to few tens of seconds.}
{In the literature, there is no attempt to understand in which order these classes 
were exhibited. Since the observation was not continuous, the appearances of these classes
seem to be in random order. Our goal is to find a natural sequence of these classes
and compare with the existing observations.
We also wish to present a physical interpretation of the sequence 
so obtained using two component advective flow model of black hole accretion.} 
{In the present paper, we compute the ratios of the power-law photons and the black  
body photons in the spectrum of each class and call these ratios as the `Comptonizing efficiency' (CE). 
We sequence the classes from the low to the high value of CE. The number of photons were
obtained by fitting the spectra of two independent sets of data of each class with disk 
blackbody and power-law components, after making suitable correction for the absorption in the 
intervening medium.}
{We clearly find that each variability class could be characterized by a unique average 
Comptonizing efficiency. The sequence of the classes based on this parameter
seem to be corroborated by a handful of the observed transitions caught
by Rossi X-ray timing explorer and the Indian payload Indian X-ray Astronomy Experiment
and we believe that future observation of the object would show 
that the transitions can only take place between consecutive classes in this 
sequence. Since the power-law photons are produced by inverse Comptonization of the
intercepted soft-photons from the Keplerian disk, a change in CE actually corresponds to 
a change in geometry of the Compton cloud. Thus we claim that the size of the Compton cloud 
gradually rises from very soft class to the very hard class.}
{}
\maketitle
%

\section{Introduction}

The enigmatic stellar mass black hole binary GRS 1915+105 (Harlaftis \& Greiner, 2004) 
was first discovered in 1992 by the WATCH detectors (Castro-Tirado et al. 1992) as a transient source 
with a significant variability in X-ray photon counts (Castro-Tirado et al. 1994). 
In the RXTE era, GRS 1915+105 was monitored thousands of times in the X-ray band and 
the scientific results reveal a unique nature of this compact object. The radio 
observation with VLA suggests apparent superluminal nature of its radio jets.
Radio observation  constrains that its maximum distance is 
no more than $13.5$ kpc and that the jet axis makes an angle of $70^{\circ}$ 
with the line of sight (Mirabel \& Rodriguez, 1994). 

Continuous X-ray observation of GRS 1915+105 reveals that the X-ray intensity of the source changes 
peculiarly in a variety of timescales ranging from seconds to days (Greiner et al., 1996, Morgan 
et al., 1997). Quasi-Periodic Oscillations (QPOs) are observed in a wide range of frequencies. 
QPOs in this source are associated with different types of X-ray variabilities and their timing properties 
are correlated with spectral features (Muno et al., 1999, Sobczak et al., 1999, Rodriguez et al., 2002,
 Vignarca et al., 2003,).  The origin of QPO frequencies between 
$0.5$ to $10$ Hz is identified to be due to the oscillation of the Comptonized 
photons, presumably emitted from the post-shock region of the low angular momentum (sub-Keplerian)
flow (Chakrabarti \& Manickam, 2000, hereafter CM00; Rao et al., 2000). 

Small scale variabilities of GRS 1915+105 are identified with local variation of the inner disk 
(Nandi et al., 2000, Chakrabarti \& Manickam, 2000; 
Migliari \& Belloni, 2003). Several observers have reported that 
this object exhibited many types of variability classes 
(Yadav et al., 1999; Rao, Yadav \& Paul, 2000; Belloni et al. 2000, 
Chakrabarti \& Nandi, 2000; Naik et al. 2002a). Depending on the variation 
of photon counts in different arbitrary energy bands (hardness ratio) and color-color diagram of 
GRS 1915+105, the X-ray variability of the source was found to have
fifteen arbitrarily named ($\alpha,\ \beta,\ \gamma,\ \delta,\ \phi,\ 
\chi_1, \chi_2, \chi_3, \chi_4,\ \mu,\ \nu,\ \lambda,\ \kappa,\ \rho,\ \theta$)
classes. In a 1999 observation of RXTE, the existence of another class $\omega$ was reported
(Klein-Wolt et al., 2002; Naik et al. 2002a). In the so-called $\chi$ (i.e., $\chi_1$ to $\chi_4$) 
class, the strong variability as is found in other classes is absent. 
The classes named $\chi_1, \chi_3$, $\beta$ and $\theta$ are 
associated with the presence of strong radio jets (Naik \& Rao, 2000, Vadawale et al., 2003). 
To understand the above features from a dynamical point of view, we carried out a correlation 
study in between temporal and spectral features of this source in different classes. A preliminary 
report is presented in Pal, Nandi \& Chakrabarti (2008, hereafter PNC08). 

While a large number of papers have been published in the literature on GRS 1915+105,
to our knowledge, there is no work which actually asked the question: are these classes arbitrary,
or they appear in a given sequence? The problem lies in the fact that no satellite 
continuously observed GRS 1915+105. Sporadic observations caught the object in sporadic
classes. In the present paper, we try to show that these classes could be parameterized by  a
common parameter, namely, the ratio between the number of photons 
in the power-law component and the black body component. We call this as the 
Comptonizing Efficiency or CE. Since the number of 
photons in the power-law component depends on the degree of interception by the 
so-called hot electron cloud or Compton cloud (Sunyaev \& Titarchuk, 1980, 1985), different 
classes are therefore parameterized by the average size of the Compton cloud. Along with the
dynamical evolution of CE, we compute the spectrum and the power density spectrum (PDS)
for each of the classes. To accomplish the 
computation of CE, we separate out the photons $\gamma_{BB}$ of the black body component 
and the photons $\gamma_{PL}$ from the power-law component and take the running ratio CE 
as a function of time to study how the Compton cloud itself varies in a short time 
scale. Our findings reveal that the Compton cloud is highly dynamic. We present possible 
scenarios of what might be occurring to it in different variability classes.  

The paper is organized as follows: in the next Section, we present a general discussion of 
the observation, our criteria of selection of data for analysis and analysis technique. 
In \S 3, we discuss the procedure of calculation that is adopted to calculate the photon numbers.
In \S 4, the results are presented. In \S 5, we present a unifying view where we show that the 
Comptonizing efficiency may be a key factor to distinguish among various 
classes. Finally, in \S 6, we make concluding remarks.

\section{Observation \& Data Analysis}

The RXTE science data is taken from the NASA HEASARC data archive for analysis. We have chosen
the data procured in 1996-97 by RXTE as in this period GRS 1915+105 has shown almost
all types of variabilities in X-rays. Subsequently, in 1999, another class, namely $\omega$ was
seen. In the present paper, since we are interested in sequencing these sixteen classes,
we will rename them as follows: I=$\phi$; II=$\delta$; III=$\gamma$; IV=$\omega$; V=$\mu$;
VI=$\nu$; VII=$\lambda$; VIII=$\kappa$; IX=$\rho$; X=$\beta$; XI=$\alpha$; XII=$\theta$; 
XIII=$\chi_2$; XIV=$\chi_4$; XV=$\chi_1$; XVI=$\chi_3$). This would be our final sequence
also, and hence it is easier to remember. Moreover, theoretical discussions often uses
these Greek characters it would be confusing to use these symbols.

During the data analysis we excluded the data collected for elevation angles less than $10^{\circ}$,
offset greater than $0.02^{\circ}$ and during the South Atlantic Anomaly (SAA) passage. In Table. 1, the
details of the data selection and ObsIDs are given which we analyzed in this paper.

\begin{table}
\centering
\begin{tabular}{ccc}
Obs-Id & Class & Date \\
\hline\hline
10408-01-19-00$^*$ & I &29-06-1996\\
10408-01-09-00 & I &29-05-1996\\
10408-01-18-00$^*$ &II &25-06-1996\\
20402-01-41-00 & II &19-08-1997\\
20402-01-37-00$^*$ & III &17-07-1997\\
20402-01-56-00 & III &22-11-1997\\
40703-01-27-00$^*$ & IV &23-08-1999\\
40403-01-07-00 & IV &23-04-1999\\
10408-01-36-00$^*$ & V &28-09-1996\\
20402-01-53-01 & V &05-11-1997\\
10408-01-40-00$^*$ & VI &13-10-1996\\
20402-01-02-02 & VI &14-11-1996\\
20402-01-36-00$^*$ & VII &10-07-1997\\
10408-01-38-00 & VII &07-10-1996\\
20402-01-35-00$^*$ & VIII &07-07-1997\\
20402-01-33-00 & VIII &18-06-1997\\
20402-01-31-00$^*$ & IX &03-06-1997\\
20402-01-03-00 & IX &19-11-1996\\
10408-01-10-00$^*$ & X &26-05-1996\\
20402-01-44-00 & X &31-08-1997\\
20187-02-01-00$^*$ & XI &07-05-1997\\
20402-01-30-01 & XI &28-05-1997\\
10408-01-15-00$^*$ & XII &16-06-1996\\
20402-01-45-02 & XII &05-09-1997\\
20402-01-16-00$^*$ & XIII &22-02-1997\\
20402-01-05-00 & XIII &04-12-1996\\
20402-01-25-00$^*$ & XIV &19-04-1997\\
10408-01-33-00 & XIV &07-09-1996\\
10408-01-23-00$^*$ & XV &14-07-1996\\
10408-01-30-00 & XV &18-08-1996\\
20402-01-50-00$^*$ & XVI &14-10-1997\\
20402-01-51-00 & XVI &22-10-1997\\
\end{tabular}
\caption{The ObsIDs and the dates of RXTE data analyzed in this paper. $^*$ represents the result shown 
in Fig.~3 of this paper.}
\label{table:1}      
\end{table}
 
Here we discuss how we analyzed the temporal and spectral properties of the data.

\subsection{Timing Analysis}
In timing analysis of the RXTE/PCA data we use ``binned mode'' data which was available
for 0-35 channels only, with a time resolution of $2^{-7}$ sec and `event mode' data with a time
resolution of $2^{-8}$ sec for the rest of the channels. We restrict ourselves in the energy
range of $\sim 2-40$ keV for the timing analysis of PCA data. After extracting the light
curves (using standard tasks) from the two sets of different modes, we add them by the FTOOLS
task ``lcmath'' to have the whole energy range light curve ($2$ to $40$ keV). The Power Density
Spectrum (PDS) is generated by the standard FTOOLS task ``powspec'' with normalization.
Data is re-binned to $0.01$s time resolution to obtain a Nyquist frequency of $50$ Hz as the power
beyond this is found to be insignificant. PDSs are normalized to give the squared {\it rms}
fractional variability per Hertz.

To have the timing evolution of QPO features in different classes, we generate PCA
light curves of $2$ to $40$keV photons with $0.01$s time bin. We use the science data
accordingly from the `binned mode' and the `event mode' for $10$s time interval in each step.
This light curve is then used to make PDS using the ¨powspec¨ task. The dynamic PDS is made
by considering the shift of time interval by $1$s. The selection of time interval is done by
FTOOLS task `timetrans'. The PDSs are plotted accordingly to see the variation of QPO features
in a particular class of a few hundred seconds of observation.

\subsection{Spectral Analysis}

Spectral analysis for the PCA data is done by using ``standard2'' mode data which
have $16$ sec time resolution and we constrained our energy selection up to $40$ keV to match with
the timing analysis. The source spectrum is generated using FTOOLS task ``SAEXTRCT'' with $16$ sec
time bin from ``standard2'' data.
The background fits file is generated from the ``standard2'' fits file by the
FTOOLS task ``runpcabackest'' with the standard FILTER file provided with the package.
The background source spectrum is generated using FTOOLS task ``SAEXTRCT'' with $16$ sec time bin from
background fits file. The standard FTOOLS task ``pcarsp'' is used to generate 
the response file with appropriate detector information.
The spectral analysis and modeling was performed using XSPEC (v.12) astrophysical fitting
package. For the model fitting of PCA spectra, we have used a systematic error of $1\%$. The
spectra are fitted with diskbb and power law model along with $6.0\times 10^{22} cm^{-2}$ hydrogen
column absorption (Muno et al., 1999) and we used the Gaussian for iron line as required for best fitting.
During fitting of the spectra we have adopted the method used by Sobczak et al., (1999), 
to obtain the values of spectral parameters. We have calculated error-bars of 90\% confidence 
level in each case. In case of soft states, the diskbb is found be fitted within $2.5-10.0$ keV. But in case 
of hard states, the diskbb is fitted within $2.5-5.0$ keV. This upper limit changes dynamically 
for the fitting of spectra. Table.~2 gives the energy up to which the black body fit is made for each class.

To have the spectral evolution with time for each class, we have generated PCA spectrum
($2.5$ to $40$keV) with a minimum of $16$s time interval along with background spectrum and response
matrix. This procedure is repeated with every $16$s shift in time interval since the minimum
time resolution in 'standard2' data is $16$s. We use `timetrans' task to select
each and every step of time interval. Spectral evolution (dynamic representation) for a 
longer duration of observation ($\sim 500$ to $2000$ sec) of each class is plotted to show 
the local spectral variation in each class. 

\section{Calculation Procedure}

In this section, we discuss the process of calculation of photon numbers using the parameters obtained
from spectral fit.

Each spectrum is first fitted within XSPEC environment with the models and constraints mentioned above.
The fitting parameters are used to calculate the black body photon counts from the Keplerian component and
the power-law photon counts from the Compton cloud. The number of black body photons
are obtained from the fitted parameters of the multi-color disk black body model (Makishima et al., 
1986). This is given by,

\begin{equation}
f(E)=\frac{8\pi}{3} r_{in}^2 \cos{i} \int_{T_{out}}^{T_{in}} (T/T_{in})^{-11/3}B(E,T)\,dT/T_{in},
\end{equation}
where, $B(E,T)=\frac{E^3}{(\exp{E/T}-1)}$ and $r_{in}$ can be calculated from,

\begin{equation}
K=(r_{in}/(D/10kpc))^2 \cos{i},
\end{equation}
where, $K$ is the normalization of the blackbody spectrum obtained after fitting, $r_{in}$ is the
inner radius of the accretion disk in km, $T_{in}$ is the temperature at $r_{in}$ in keV. $D$ is
the distance of the compact object in kpc and $i$ is the inclination angle of the accretion
disk. Here, both the energy and the temperature are in keV. The black body flux, $f(E)$ in
counts/s/keV is integrated between $0.1$ keV to the maximum energy (obtained from 
diskbb model fits with the spectra as done in Sobczak et al., 1999). This
gives us with $\gamma_{BB}$, the number of the black body photons in counts/s.

The Comptonized photons $\gamma_{PL}$ that are produced due to inverse-Comptonization
of the soft black body photons by `hot' electrons in the Compton cloud are calculated by fitting with the
power-law given below,
\begin{equation}
P(E)=N(E)^{-\alpha},
\end{equation}
where, $\alpha$ is the power-law index and $N$ is the total counts/s/keV at $1$keV.
It is reported in Titarchuk (1994), that the Comptonization spectrum will have a peak at around
$3 \times T_{in}$. The power law equation is integrated from $3 \times T_{in}$ to $40$keV to
calculate total number of Comptonized photons in counts/s.
The Gaussian function  $Ga(E)$ incorporated with power law equation for the presence of
line emissions in the spectrum, whenever necessary is given below,
\begin{equation}
Ga(E)= L \frac{1}{\sigma \sqrt{2\pi}} \exp(-0.5\left(E-E_l)/\sigma)^2)\right),
\end{equation}
where, $E_l$ is the energy of emitted line in keV, $\sigma$ is the line width in keV and $L$ is the
total number of counts/s in the line. The variation of the Comptonizing Efficiency (CE) with
time is plotted to have an idea of how the Compton cloud may be changing its geometry.

We fitted the data with another Comptonization model, namely, the {\it compst}
(Sunyaev \& Titarchuk, 1980). The spectrum is given by,
\begin{eqnarray}
Cst(x)&=& \frac{\alpha(\alpha+3)}{2\alpha+3} \left( \frac{x}{x_0} \right)^{\alpha+3}, when~ 0\leq x \leq x_0, \nonumber\\
      &=& \frac{\alpha(\alpha+3){x_0}^\alpha}{\Gamma(2\alpha+4)}x^3 \exp(-x) I(x), when ~ x \geq x_0. \
\end{eqnarray}
Here $I(x)= \int^{\infty}_0 t^{\alpha-1} \exp{-t} \left( 1+\frac{t}{x} \right)^{\alpha+3},$ 
$x=E/kT_e,$ $x_0=kT_{in}/kT_e,$ and $\alpha$ is the spectral index. This $\alpha$ is calculated
from $\alpha=\sqrt{\frac{9}{4}+\gamma}-\frac{3}{2}$, where $\gamma = \frac{\pi^2 m_e c^2}
{3(\tau+\frac{2}{3})^2 kT_e}$. $kT_e$ is the electron temperature and $\tau$ is the optical depth of the
electron cloud. In the computation, we exclude the hydrogen column absorption feature while calculating CE.
This is because we are interested in the photons which were emitted from the disk before
suffering any absorption. Thus the variation of CE along with other features (e.g., spectral
and QPO frequency variations) will reflect the {\it actual} radiative properties of the flow near a black hole.  

\begin{figure*}
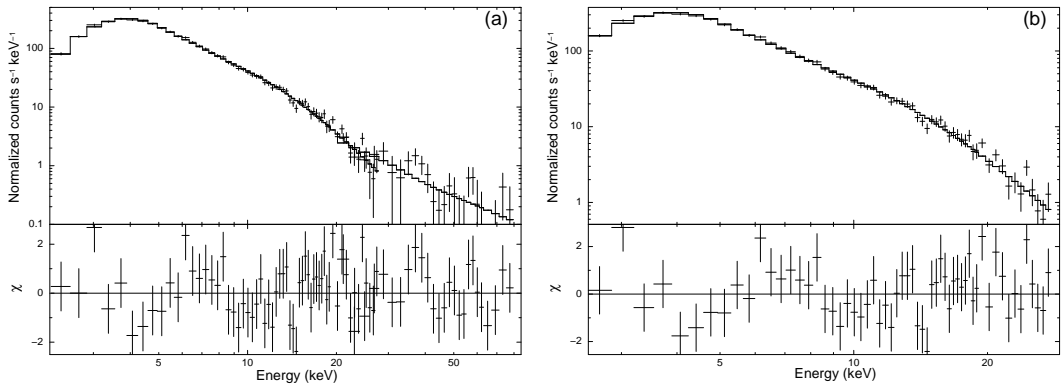

\centering
\includegraphics[angle=-90,width=7cm]{fig1a.ps}
\includegraphics[angle=-90,width=7cm]{fig1b.ps}
\caption{In the left panel, a sample fitted PCA spectrum along with the {\it diskbb}
and power law component are shown. The HEXTE data is also added.
In the right panel, the same PCA spectrum is fitted with {\it diskbb} and {\it compST} components. 
\label{compare}}
     \end{figure*}

In Fig. 1, we show an example of the fitted PCA and HEXTE spectrum along with the fitted components. 
We also show the residuals to characterize the goodness of the fit.
In the left panel, the fitting is done with {\it diskbb} and power-law components. In the right panel,
the same spectrum is fitted with {\it diskbb} and {\it compST} models.
The photon numbers and CE are calculated with the parameters obtained from the fitting.
The black body spectrum is simulated with $T_{in}=1.15~^{+0.057}_{-0.052}$ keV. The calculated number of 
black body photons from $0.1-6.5$keV is $156.03~^{+17.10}_{-13.90}$ kcounts/s. In the same way, 
the power-law spectrum is simulated with the power-law index = $2.057~^{+0.154}_{-0.147}$.
The calculated number of the Comptonized photons between $3.45-40$ keV is $0.39~^{+0.06}_{-0.05}$ kcounts/s. 
The ratio between the power-law photon and the black body photon is $0.25~^{+0.08}_{-0.07}$\%. 
This means that only $0.41$\% of the soft photons are Comptonized by the Compton cloud.
For the sake of comparison, the simulation with the {\it compST} model parameters
are found to be $KT_e = 6.41~^{+2.55}_{-1.25}$ and $\tau = 9.42~^{+2.88}_{-2.39}$. 
The number of Comptonized photons within the same range as previous case is $0.41~^{+0.06}_{-0.05}$ kcounts/s. 
Thus the ratio between the {\it compST} photon and the {\it diskbb} photon is $0.26~^{+0.08}_{-0.07}$\%. 
In Fig. 1a, we also include HEXTE data, and fitted with a power-law. But the spectral fit does not change 
in slope and the number of photons contributed by HEXTE regime is so low that the basic result of 
CE is not changed. In Fig. 1b, we give an example where {\it comptST} model cannot be fitted with 
HEXTE data. This is true for many variability classes. We therefore ignore the HEXTE data 
in the rest of the paper. 

In Table.~2, a comparison of the fitted parameters for all the classes is provided with 
error at $90\%$ confidence level. 
The first column gives the class name and the second column gives the burst-off (harder) and 
the burst-on (softer) states, if present, in each class.  These states are actually similar to
hard and soft states of a black hole, except that the wind plays a major role
and thus the states are not all the way hard or soft. The third column gives the black body temperature
of the fitted Keplerian disk ($T_{in}$ in keV). The fourth column gives the maximum energy in keV upto 
which {\it diskbb} model is fitted and soft photon number is calculated. The fifth column
gives the count rates of the soft photons. The next column gives the index associated with the 
fitted power-law component. The seventh column gives the derived hard photon counts per second. 
The eighth column gives the ratio of the photon numbers given in the fifth and the seventh columns. This 
is the so-called Comptonizing efficiency or CE. We arranged the classes in a way that the average 
CE changes monotonically. We find it to be high in harder classes such as XIII-XVI and low
in the softer classes such as I-II. The final column gives the value of reduced $\chi^2$ for the fits. 
The behavior of average CE will be discussed later.

As mentioned earlier, we fit the spectrum of each class with {\it diskbb} and {\it compST} model
to have the information of $\tau$ and $T_e$ of electron distribution inside the Compton cloud. 
One sample spectrum is taken for each class and fitted with the {\it diskbb} and {\it compST }
model with $n_H=6.0 \times 10^{22} cm^{-2}$ hydrogen column for absorption and $1\%$ systematic error.
The best fitted parameters are given in Table.~3 for comparison with the parameters given in Table.~2. 
We find that in the cases where both models were used, the CE obtained were very close to each other.

\begin{table*}
\begin{center}
\begin{tabular}{|c|c|c|c|c|c|c|c|c|}
\hline
\multicolumn{2}{|c|}{} & $T_{in}$ & Energy & Soft & Power law &  Hard & CE  & \\
\multicolumn{2}{|c|}{Class} &  &  &  Photon & index &  Photon &  &$\chi^2$ \\
\multicolumn{2}{|c|}{} & (keV) & keV &  (kcnt/s) &   & (kcnt/s) & (\%) & \\
\hline
I &-&$1.67~^{+0.017}_{-0.017}$& 10.2 &$275.08~^{+8.77}_{-8.46}$ & $2.927~^{+0.266}_{-0.255}$ &$0.20~^{+0.04}_{-0.03}$ & $0.07~^{+0.02}_{-0.01}$ & 1.6\\ 
\hline
II &-&$1.65~^{+0.016}_{-0.015}$& 10.3 &$330.40~^{+9.48}_{-9.80}$ & $2.348~^{+0.238}_{-0.228}$ &$0.16~^{+0.03}_{-0.02}$ & $0.05~^{+0.01}_{-0.01}$ & 1.1\\ 
\hline
III &-&$1.80~^{+0.015}_{-0.014}$& 10.4 &$511.13~^{+12.57}_{-12.22}$ & $2.916~^{+0.352}_{-0.309}$ & $0.31~^{+0.09}_{-0.07}$ & $0.06~^{+0.02}_{-0.01}$ & 0.60\\ 
\hline
IV &-&$1.937~^{+0.025}_{-0.024}$& 10.6 &$192.21~^{+7.80}_{-7.41}$ & $2.649~^{+0.308}_{-0.292}$ &$0.09~^{+0.02}_{-0.01}$ & $0.05~^{+0.01}_{-0.01}$ & 0.98  \\ 
\hline
V &-&$1.421~^{+0.023}_{-0.022}$& 7.0 &$468.53~^{+10.40}_{-10.02}$ & $2.330~^{+0.145}_{-0.140}$ &$0.63~^{+0.085}_{-0.073}$ & $0.13~^{+0.02}_{-0.02}$ & 1.01\\ 
\hline
VI &-&$1.752~^{+0.030}_{-0.029}$& 7.0 &$370.39~^{+14.15}_{-13.15}$ & $2.578~^{+0.123}_{-0.119}$ &$0.45~^{+0.04}_{-0.03}$ & $0.12~^{+0.016}_{-0.015}$ & 1.18\\ 
\hline
VII &h& $0.933~^{+0.056}_{-0.049}$& 4.5 &$229.10~^{+28.94}_{-22.32}$ & $2.106~^{+0.116}_{-0.112}$ &$0.59~^{+0.07}_{-0.06}$ & $0.26~^{+0.08}_{-0.07}$ & 0.95 \\ 
\cline{2-9}
          &s& $1.992~^{+0.022}_{-0.021}$& 9.0 &$437.22~^{+11.11}_{-10.62}$ & $2.683~^{+0.219}_{-0.211}$ &$0.19~^{+0.3}_{-0.2}$ & $0.04~^{+0.007}_{-0.006}$ & 1.2 \\ 
\hline
VIII &h& $1.150~^{+0.057}_{-0.052}$& 6.5 & $156.03~^{+17.10}_{-13.90}$ & $2.057~^{+0.154}_{-0.147}$ &$0.39~^{+0.06}_{-0.05}$ & $0.25~^{+0.08}_{-0.07}$ & 1.2  \\ 
\cline{2-9}
          &s& $1.900~^{+0.027}_{-0.026}$& 8.0 & $432.72~^{+13.95}_{-13.15}$ & $2.983~^{+0.243}_{-0.234}$ &$0.21~^{+0.03}_{-0.03}$ & $0.05~^{+0.010}_{-0.008}$ & 1.5  \\
\hline
IX &h&$1.382~^{+0.031}_{-0.029}$& 5.75 &$382.33~^{+18.42}_{-16.73}$ & $2.505~^{+0.101}_{-0.099}$ &$0.95~^{+0.09}_{-0.08}$ & $0.25~^{+0.04}_{-0.03}$ & 1.54 \\ 
\cline{2-9}
       &s&$1.843~^{+0.022}_{-0.022}$& 8.0 &$419.92~^{+11.55}_{-10.97}$ & $2.331~^{+0.160}_{-0.155}$ &$0.34~^{+0.04}_{-0.03}$ & $0.08~^{+0.01}_{-0.01}$ & 0.99  \\ 
\hline
X &h& $1.042~^{+0.032}_{-0.030}$& 5.25 &$320.15~^{+20.78}_{-18.15}$ & $2.142~^{+0.061}_{-0.060}$ &$0.87~^{+0.05}_{-0.04}$ & $0.27~^{+0.04}_{-0.04}$ & 1.05   \\ 
\cline{2-9}
        &s& $1.448~^{+0.023}_{-0.022}$& 7.0 &$354.87~^{+12.38}_{-11.58}$ & $2.688~^{+0.176}_{-0.170}$ &$0.32~^{+0.05}_{-0.04}$ & $0.14~^{+0.03}_{-0.02}$ & 1.5\\ 
\hline
XI &-&$1.213~^{+0.082}_{-0.070}$& 4.5  &$261.36~^{+38.46}_{-28.74}$ & $2.067~^{+0.062}_{-0.061}$ &$0.81~^{+0.04}_{-0.04}$ & $0.31~^{+0.08}_{-0.07}$ & 1.01  \\ 
\hline
XII &h&$1.391~^{+0.039}_{-0.037}$& 6.0 &$416.84~^{+25.73}_{-22.85}$ & $2.501~^{+0.089}_{-0.087}$ &$1.15~^{+0.09}_{-0.08}$ & $0.28~^{+0.05}_{-0.04}$ & 1.3  \\ 
\cline{2-9}
         &s&$1.317~^{+0.034}_{-0.032}$& 5.5 &$388.95~^{+21.46}_{-19.19}$ & $3.248~^{+0.157}_{-0.152}$ &$0.65~^{+0.08}_{-0.07}$ & $0.17~^{+0.03}_{-0.03}$ & 1.5  \\ 
\hline
XIII &-&$1.200~^{+0.117}_{-0.094}$& 4.5 &$126.80~^{+27.74}_{-18.46}$ & $1.968~^{+0.076}_{-0.074}$ &$0.57~^{+0.03}_{-0.03}$ & $0.45~^{+0.17}_{-0.12}$ & 1.2  \\ 
\hline
XIV &-&$1.184~^{+0.113}_{-0.091}$& 4.5 &$121.04~^{+25.87}_{-17.37}$ & $1.954~^{+0.081}_{-0.079}$ &$0.52~^{+0.03}_{-0.02}$ & $0.43~^{+0.16}_{-0.12}$ & 1.0  \\ 
\hline
XV &-&$1.165~^{+0.053}_{-0.048}$& 4.5 &$599.39~^{+58.18}_{-47.51}$ & $2.736~^{+0.057}_{-0.056}$ &$4.62~^{+0.12}_{-0.11}$ & $0.77~^{+0.08}_{-0.07}$ & 1.3  \\ 
\hline
XVI &-&$1.187~^{+0.030}_{-0.028}$& 5.0 &$598.83~^{+31.99}_{-28.57}$ & $2.702~^{+0.057}_{-0.056}$ &$5.30~^{+0.09}_{-0.09}$ & $0.88~^{+0.10}_{-0.09}$ & 1.1  \\ 
\hline
\end{tabular}
\end{center}
\caption{Parameters for the simulated spectral fits with {\it diskbb} plus {\it powerlaw} 
models for all the variability classes.}
\end{table*}

\begin{table*}
\begin{center}
\begin{tabular}{|c|c|c|c|c|c|c|c|c|c|}
\hline
\multicolumn{2}{|c|}{}& $T_{in}$ & Energy & Soft & $T_e$ & $\tau$ &  Hard & CE  & \\
\multicolumn{2}{|c|}{Class} &  &  &  Photon & & & Photon &  &$\chi^2$ \\
\multicolumn{2}{|c|}{}& (keV) & keV &  (kcnt/sec) & keV &  & (kcnt/s) & (\%) & \\
\hline
I &-&$1.67~^{+0.017}_{-0.017}$& 10.2 & $275.08~^{+8.77}_{-8.46}$ &-&-&-&-&- \\ 
\hline
II &-&$1.65~^{+0.016}_{-0.015}$& 10.3 &$330.04~^{+9.48}_{-9.80}$ &-&-&-&-&- \\ 
\hline
III &-&$1.80~^{+0.015}_{-0.014}$& 10.4 &$511.13~^{+12.57}_{-12.22}$ &-&-&-&-&- \\ 
\hline
IV &-&$1.937~^{+0.025}_{-0.024}$& 10.6 &$192.21~^{+7.80}_{-7.41}$ &-&-&-&-&-   \\ 
\hline
V &-&$1.421~^{+0.023}_{-0.022}$& 7.0 &$468.53~^{+10.40}_{-10.02}$ & $4.718~^{+0.983}_{-0.660}$ & $14.399~^{+4.624}_{-4.457}$ & $0.61~^{+0.085}_{-0.073}$ & $0.13~^{+0.02}_{-0.02}$ & 1.01  \\ 
\hline
VI &-&$1.752~^{+0.030}_{-0.029}$& 7.0 &$370.39~^{+14.15}_{-13.15}$ &-&-&-&-&- \\ 
\hline
VII &h& $0.933~^{+0.056}_{-0.049}$& 4.5 &$229.10~^{+28.94}_{-22.32}$ & $4.926~^{+1.082}_{-0.680}$ & $12.432~^{+2.747}_{-2.246}$ & $0.52^{+0.07}_{-0.06}$ & $0.22~^{+0.08}_{-0.07}$ & 1.5 \\ 
\cline{2-10}
          &s & $1.992~^{+0.022}_{-0.021}$& 9.0 &$437.22~^{+11.11}_{-10.62}$ &-&-&-&-&-  \\ 
\hline
VIII &h& $1.150~^{+0.057}_{-0.052}$& 6.5 & $156.03~^{+17.10}_{-13.90}$ & $6.410~^{+2.558}_{-1.255}$ & $9.415~^{+2.879}_{-2.390}$ & $0.41~^{+0.06}_{-0.05}$ & $0.26~^{+0.08}_{-0.07}$ & 0.90  \\ 
\cline{2-10}
          &s& $1.900~^{+0.027}_{-0.026}$& 8.0 & $437.72~^{+13.95}_{-13.15}$ &-&-&-&-&-   \\
\hline
IX &h&$1.382~^{+0.031}_{-0.029}$& 5.75 &$382.33~^{+18.42}_{-16.73}$ & $3.892~^{+0.458}_{-0.290}$ & $30.926~^{+4.361}_{-4.360}$ & $0.75~^{+0.09}_{-0.08}$ & $0.20~^{+0.04}_{-0.03}$ & 1.43 \\ 
\cline{2-10}
       &s&$1.843~^{+0.022}_{-0.022}$& 8.0 &$419.92~^{+11.55}_{-10.97}$ &-&-&-&-&-  \\ 
\hline
X  &h& $1.042~^{+0.032}_{-0.030}$& 5.25 &$320.15~^{+20.78}_{-18.15}$ & $4.612~^{+0.820}_{-0.563}$ & $11.075~^{+2.876}_{-2.097}$ & $0.57~^{+0.05}_{-0.04}$ & $0.18~^{+0.04}_{-0.04}$ & 0.92   \\ 
\cline{2-10}
         &s& $1.448~^{+0.023}_{-0.022}$& 7.0 &$354.87~^{+12.38}_{-11.58}$ &-&-&-&-&- \\ 
\hline
XI &-&$1.213~^{+0.082}_{-0.070}$& 4.5 &$261.36~^{+38.46}_{-28.74}$ & $5.043~^{+0.512}_{-0.410}$ & $11.906~^{+1.545}_{-1.321}$ & $0.71~^{+0.04}_{-0.04}$ & $0.27~^{+0.08}_{-0.07}$ & 1.37 \\ 
\hline
XII &h&$1.391~^{+0.039}_{-0.037}$& 6.0 &$416.84~^{+25.73}_{-22.85}$ & $4.131~^{+0.673}_{-0.472}$ & $14.412~^{+1.486}_{-4.102}$ &$0.96~^{+0.09}_{-0.08}$ & $0.43~^{+0.07}_{-0.07}$ & 0.98  \\ 
\cline{2-10}
         &s& $1.317~^{+0.034}_{-0.032}$ & 5.5 & $388.95~^{+21.46}_{-19.19}$ &-&-&-&-&-  \\ 
\hline
XIII &-&$1.200~^{+0.117}_{-0.094}$& 4.5 &$126.80~^{+27.74}_{-18.46}$ & $8.918~^{+1.952}_{-2.305}$ & $6.802~^{+1.935}_{-2.637}$ & $0.52^{+0.03}_{-0.03}$ & $0.41~^{+0.17}_{-0.12}$ & 1.0  \\ 
\hline
XIV &-&$1.184~^{+0.113}_{-0.091}$& 4.5 &$121.04~^{+25.87}_{-17.37}$ & $6.616~^{+2.744}_{-1.227}$ & $8.791~^{+2.221}_{-2.154}$ &$0.51~^{+0.03}_{-0.02}$ & $0.42~^{+0.16}_{-0.12}$ & 0.88  \\ 
\hline
XV &-&$1.165~^{+0.053}_{-0.048}$& 4.5 &$599.39~^{+58.18}_{-47.51}$ & $4.643~^{+0.823}_{-0.550}$ & $8.500~^{+1.581}_{-1.402}$ & $4.10~^{+0.12}_{-0.11}$ & $0.68~^{+0.08}_{-0.07}$ & 1.0  \\ 
\hline
XVI &-&$1.187~^{+0.030}_{-0.028}$& 5.0 &$598.83~^{+31.99}_{-28.57}$ & $6.616~^{+2.744}_{-1.227}$ & $8.791~^{+2.221}_{-2.154}$ & $5.20~^{+0.09}_{-0.09}$ & $0.87~^{+0.10}_{-0.09}$ & 0.88  \\ 
\hline
\end{tabular}
\end{center}
\caption{Parameters for the simulated spectral fits with {\it diskbb} and {\it comptST} 
model for all the variability classes. Those classes which could not be fitted with 
this model have been marked with dashes.}
\end{table*}

\section{Results}

We discuss below the main results for each class one by one. The Observational IDs are given in Table 1.
So far, we have shown that the CE, which were computed model independent way, vary 
from one variability class to another, i.e., our results did not depend on any 
specific theoretical model. However, to interpret the results, and to facilitate 
the discussions, it is instructive to keep a paradigm in mind. 
In Fig. 2(a-c), we present a cartoon diagram which is basically the two component 
advective flow (TCAF) model of Chakrabarti \& Titarchuk (1995, hereafter CT95),
suitably modified to include outflows as in Chakrabarti \& Nandi (2000) (Inclusion of 
the jets and outflows does not make our model a three component model, 
since the outflows are results of the inflow and formed from the post-shock region, which is 
the CENtrifugal pressure supported BOundary Layer, or CENBOL.)
In the Figure, the BH represents the black hole, the dark shaded
disk is the Keplerian flow and the light shaded disk is the sub-Keplerian flow. 
The CENBOL and the outflow from it intercept the soft photons from the Keplerian disks
and may be cooled down if the Keplerian rate is sufficient. 

   \begin{figure*}
   \centering
   \includegraphics[angle=0,width=7cm]{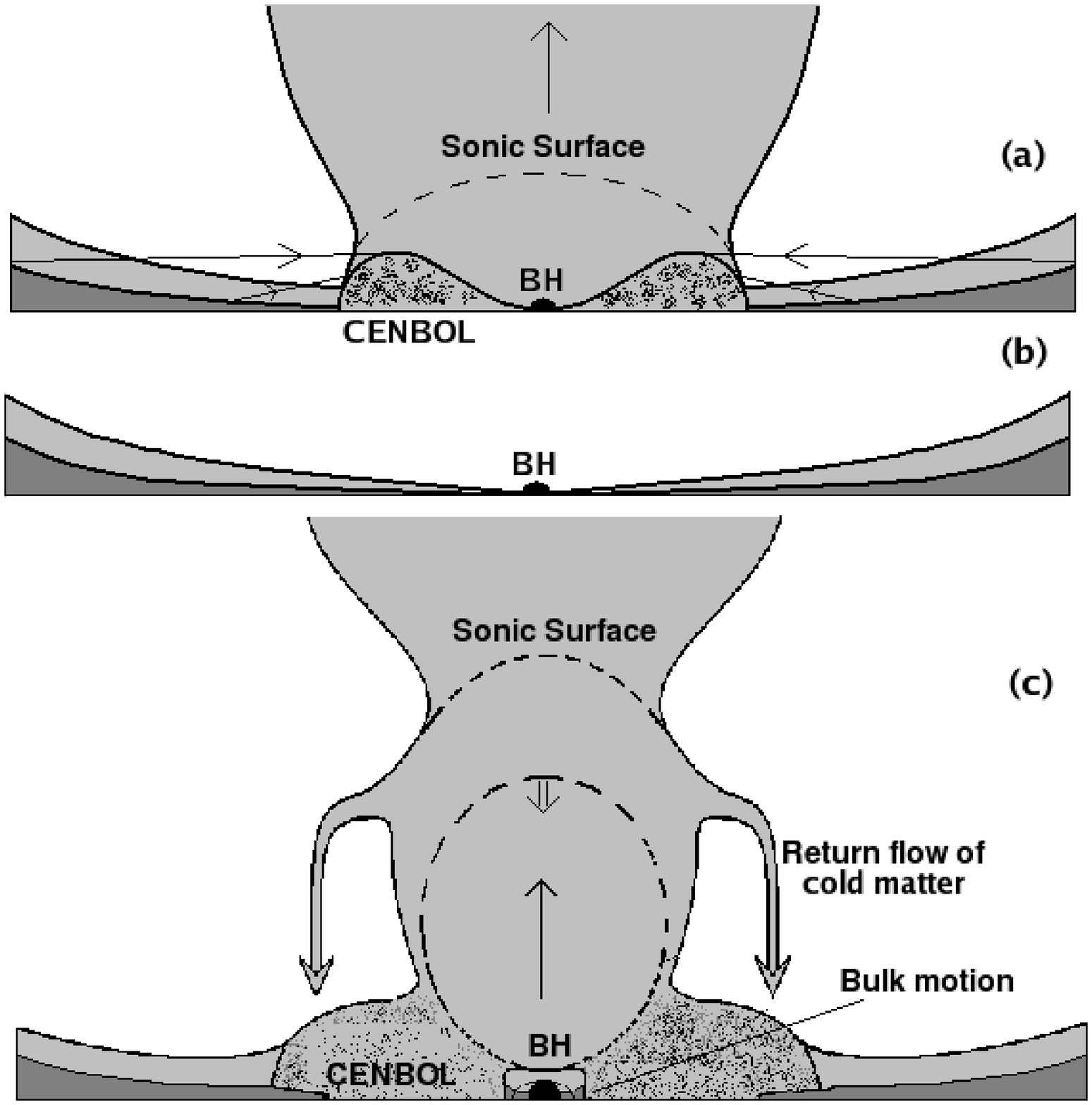}
\caption{Cartoon diagrams for three major types of variabilities in GRS 1915+105.
The classes XIII-XVI belong to the group (a) where the Keplerian rate is low, the CENBOL is large
and $<CE>$ is high. Jet/outflow rates may be low and continuous. This is the so-called `Hard' group. 
The softer classes such as I-IV belong to group (b) where the CENBOL is very small and very little 
Comptonization may be due to the sub-Keplerian flow only. This is the `Soft' group. 
The rest of the classes belong to the variations of group (c) or the `Intermediate' group, 
where, the jet also plays dynamically important role in shaping the spectrum.
\label{proof}}
\end{figure*}

In the cartoon diagram of Fig. 2a, the CENBOL is not cooled enough and the Comptonization
of the soft photon is done both by the CENBOL as well as the outflow. This configuration typically 
produces a `harder' state. However, the outflows depends on the shock strength (Chakrabarti, 1999, 
hereafter C99) and could produce weak jets as in XIII-XIV or strong jets as in 
XV and XVI (see, Vadawale et al., 2001). In Fig. 2b, the Comptonization is
due to the high accretion rate in the Keplerian disk and the CENBOL collapses. 
As a result, no jets or outflows are formed. This configuration typically produces 
the soft states. If both the Keplerian and the sub-Keplerian rates 
are comparable, then the intermediate states would be produced also. 
In Fig. 2c, we show the situation where the shock strength is intermediate and consequently, 
the outflow rate is the highest (see, C99; Chakrabarti \& Nandi, 2000 and references therein). 
In this case, there is a possibility 
that the outflow may be cooled down by Comptonization if the intercepted soft photon 
flux is high enough and the outflow is temporarily terminated. The flow falls back to 
the accretion disk, increasing local accretion rate in a very short time scale (seconds).
We believe that the variability classes (such as V-XII) some of which show clear softer (burst-on) 
and harder (burst-off) states alternately and showed evidences of intermittent 
jets (Klein-Wolt et al. 2002, Rodriguez et al. 2008) belong to this category 
(Chakrabarti \& Manickam, 2000).

We now present the dynamical analysis of the light curves of all the variability classes. 
While choosing the data duration for a given class, the following considerations have been followed:
If the count rate has no obvious rise/fall signatures, we use the data of $500$s. If the count rates
have some `repetitive' behavior, we use the data of $500$, $1000$, $1500$s or even $2000$s, which ever is 
bigger so that at least one full cycle in included in the data chunk. 
However, in the latter cases, while computing the CE, we use the data from a full cycle only. 

In each of Fig. 3(a-p) we show four panels. The top panel is the variation of the photon counts
with time. The second panel is the variation of 
log(power) of the dynamical Power Density Spectra (PDS) which may show presence or absence of quasi-periodic
variations or QPOs. The third panel is the dynamic energy spectrum which shows whether the variability class
is dominated by hard photons or soft photons. Finally, the bottom panel shows the 
variation of Comptonizing Efficiency CE calculated using $16$ seconds of binned data. The error bar is provided
at $90\%$ confidence level.

   \begin{figure*}
   \centering
{
   \includegraphics[angle=270,width=6.7cm]{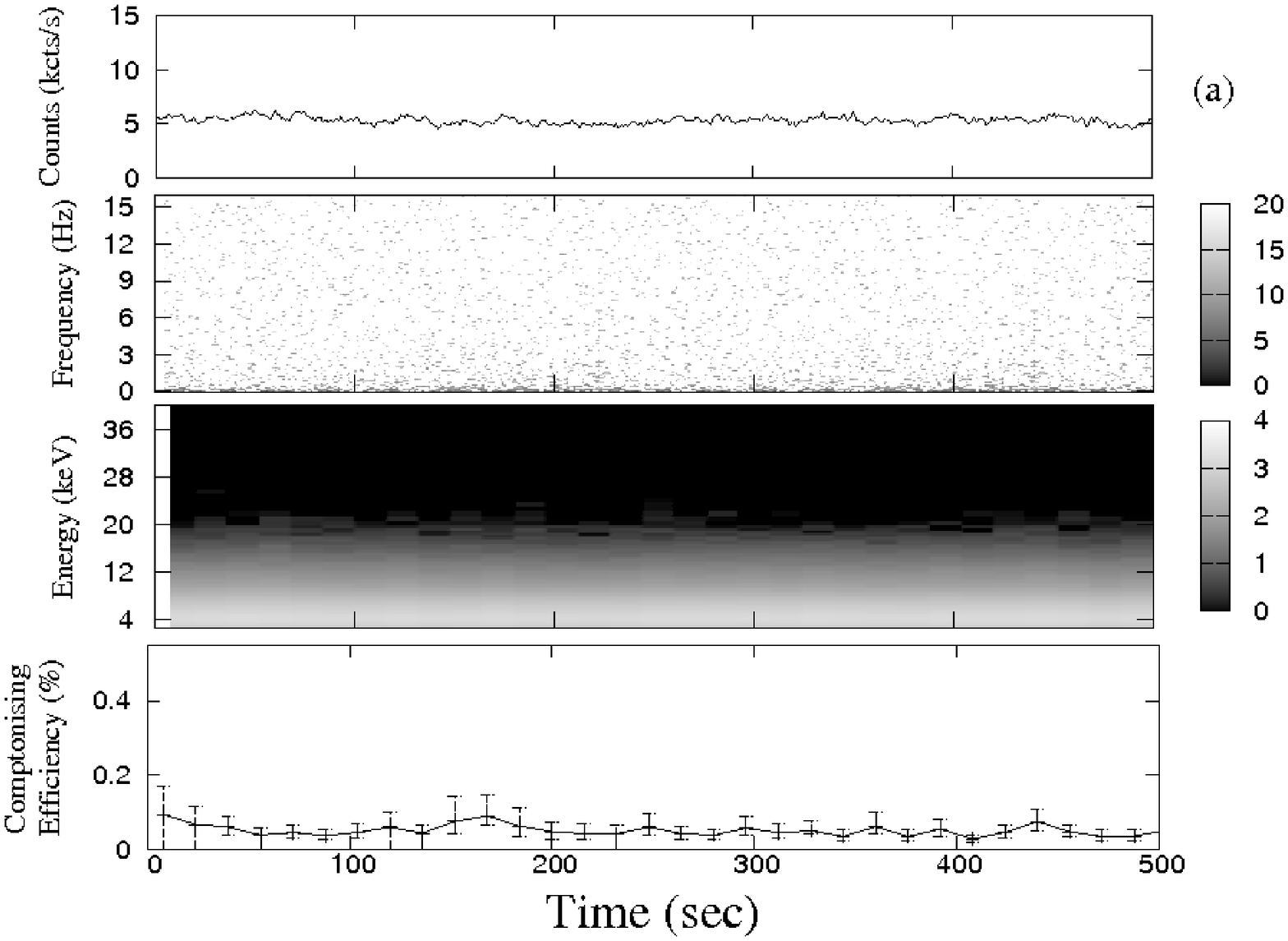}
   \includegraphics[angle=270,width=6.7cm]{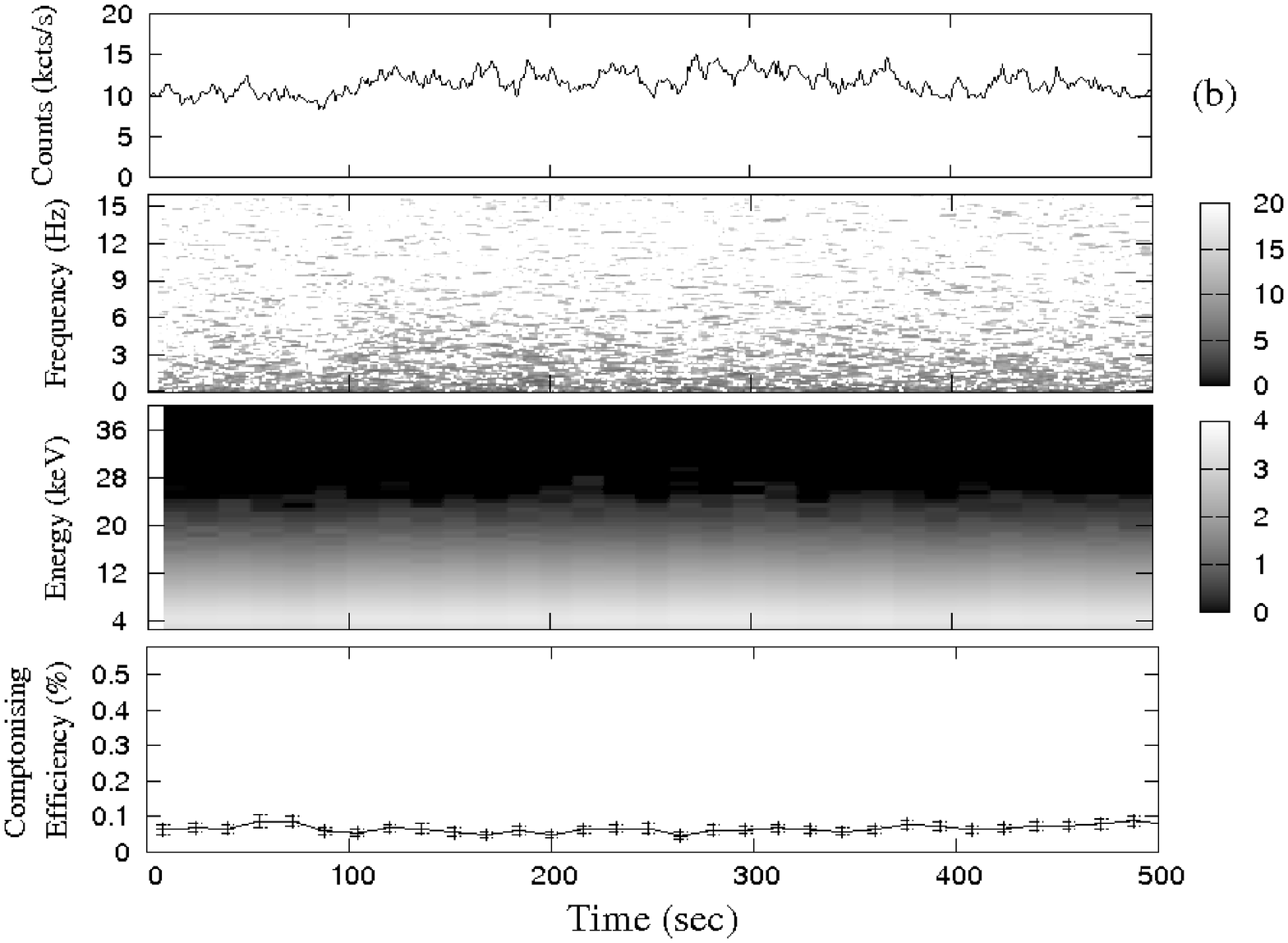}\\ 
}
   \centering
{
   \includegraphics[angle=270,width=6.7cm]{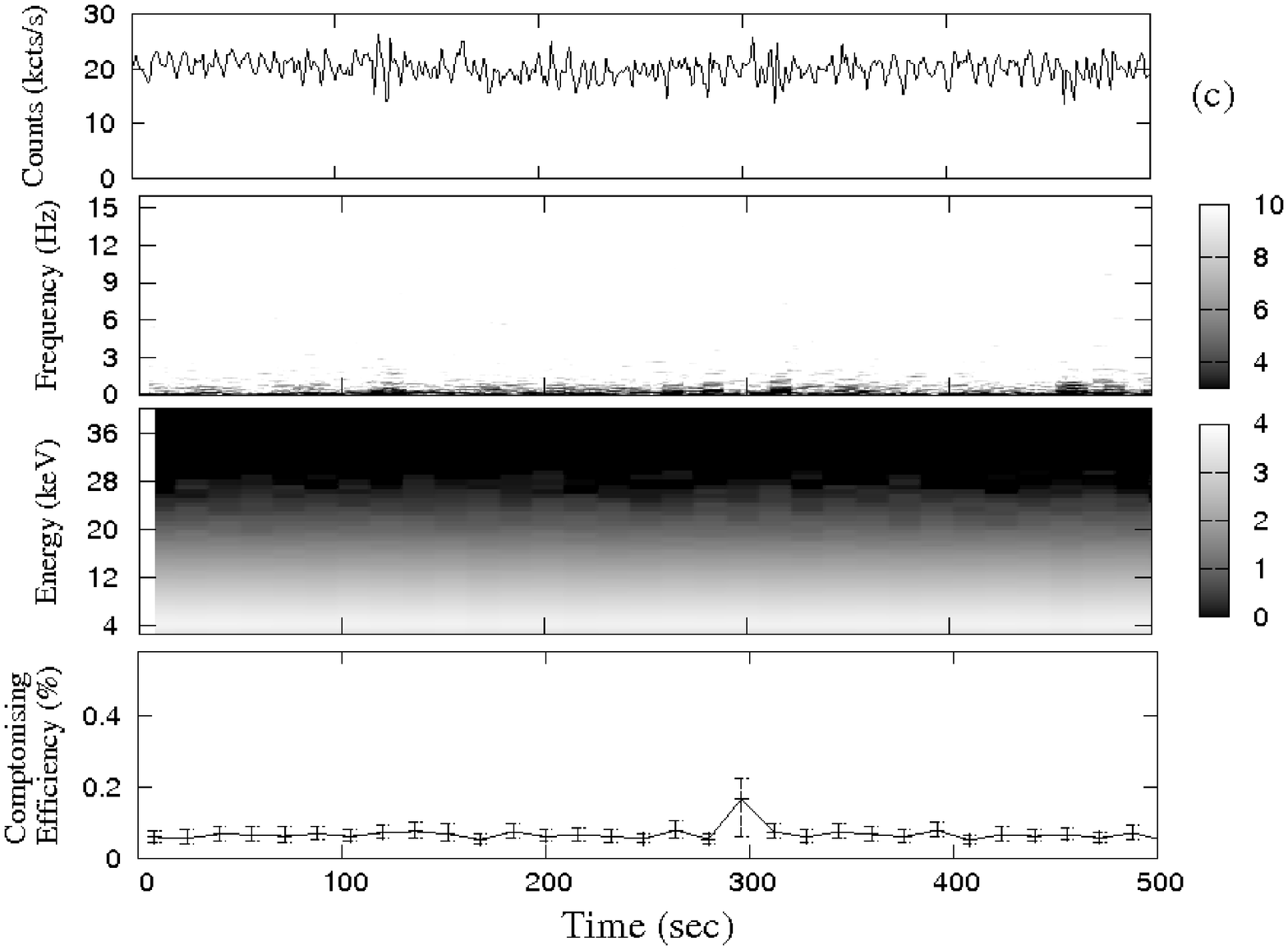}
   \includegraphics[angle=270,width=6.7cm]{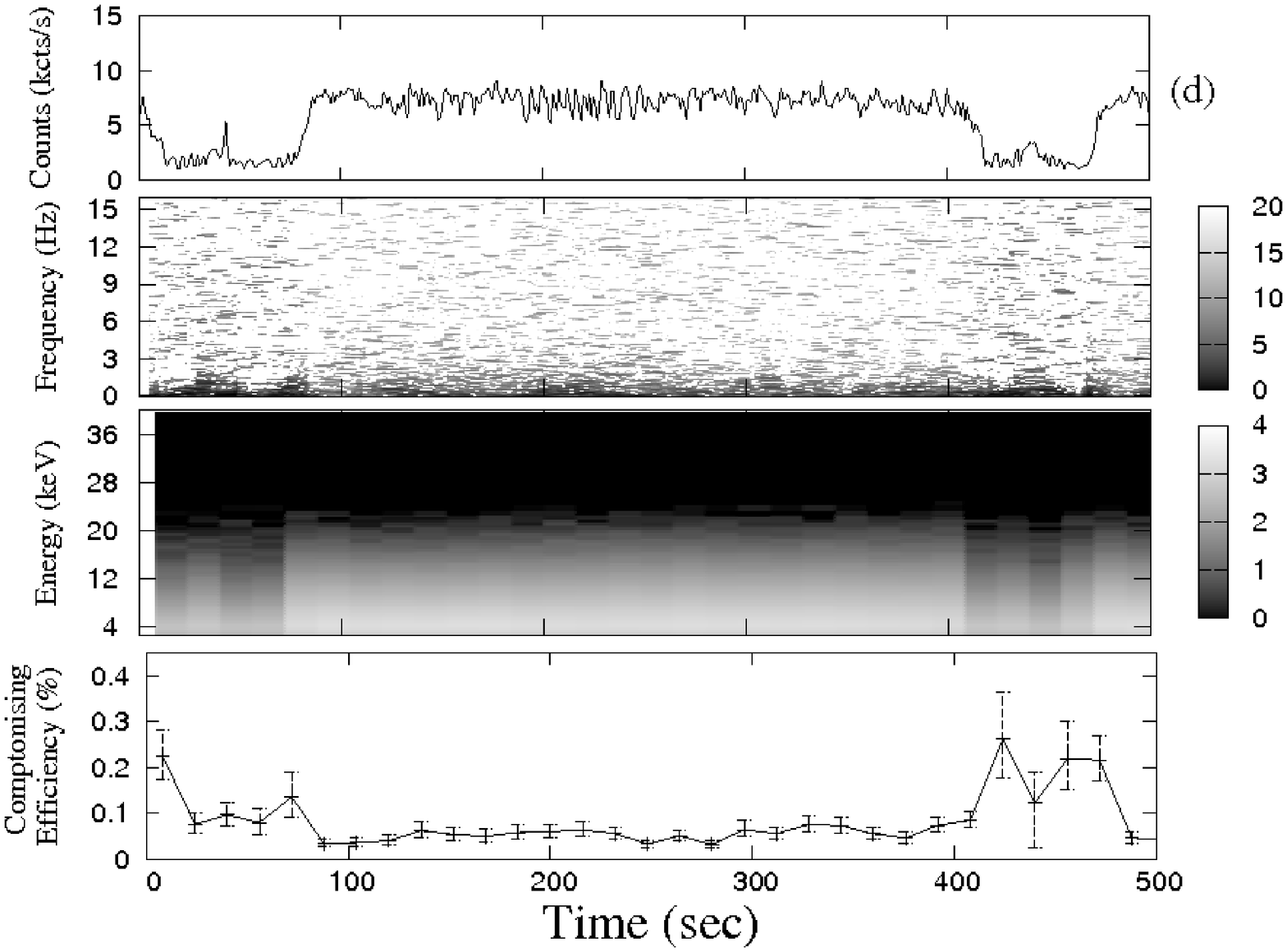}\\
}
   \centering
{
   \includegraphics[angle=270,width=6.7cm]{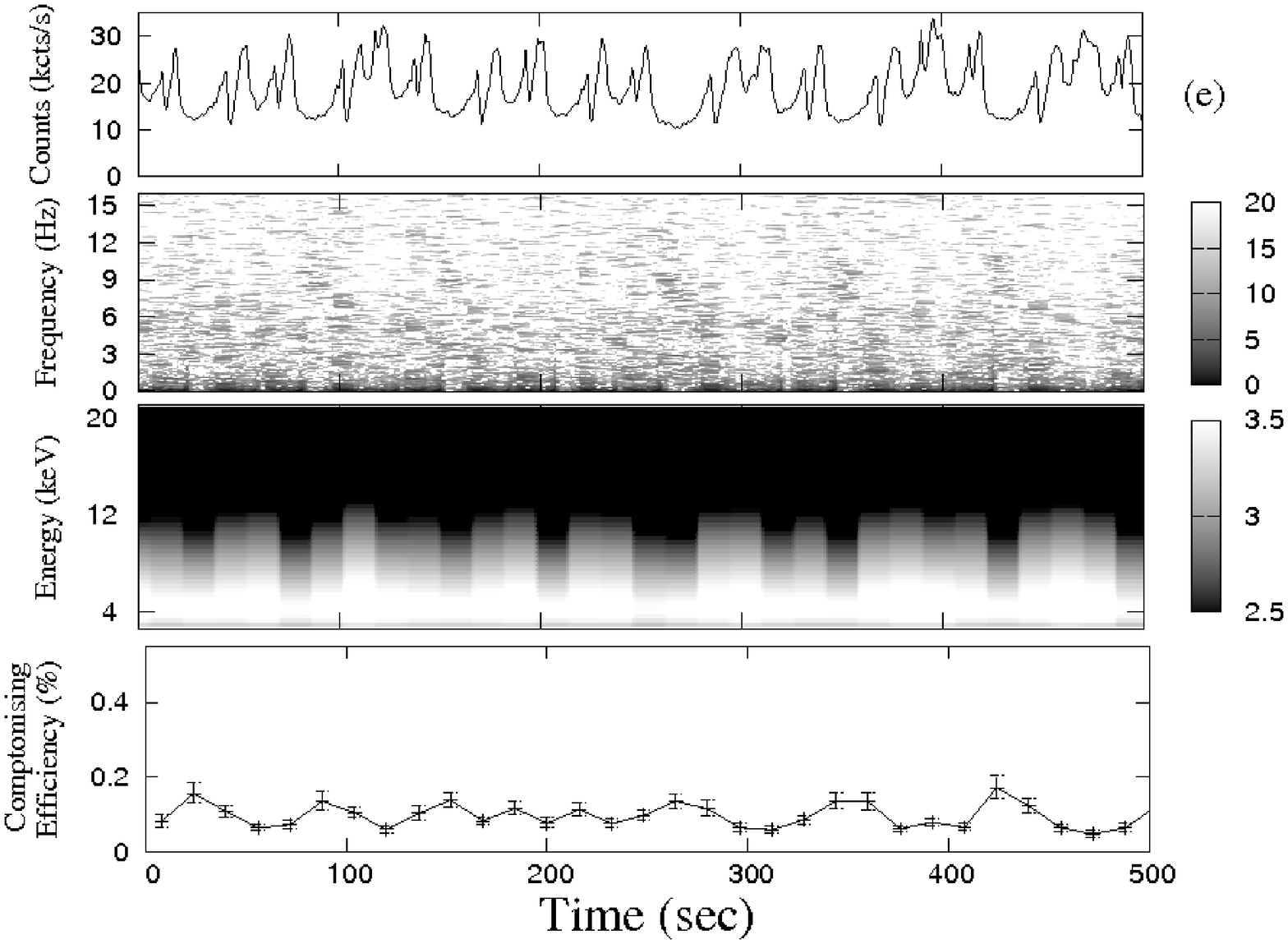}
   \includegraphics[angle=270,width=6.7cm]{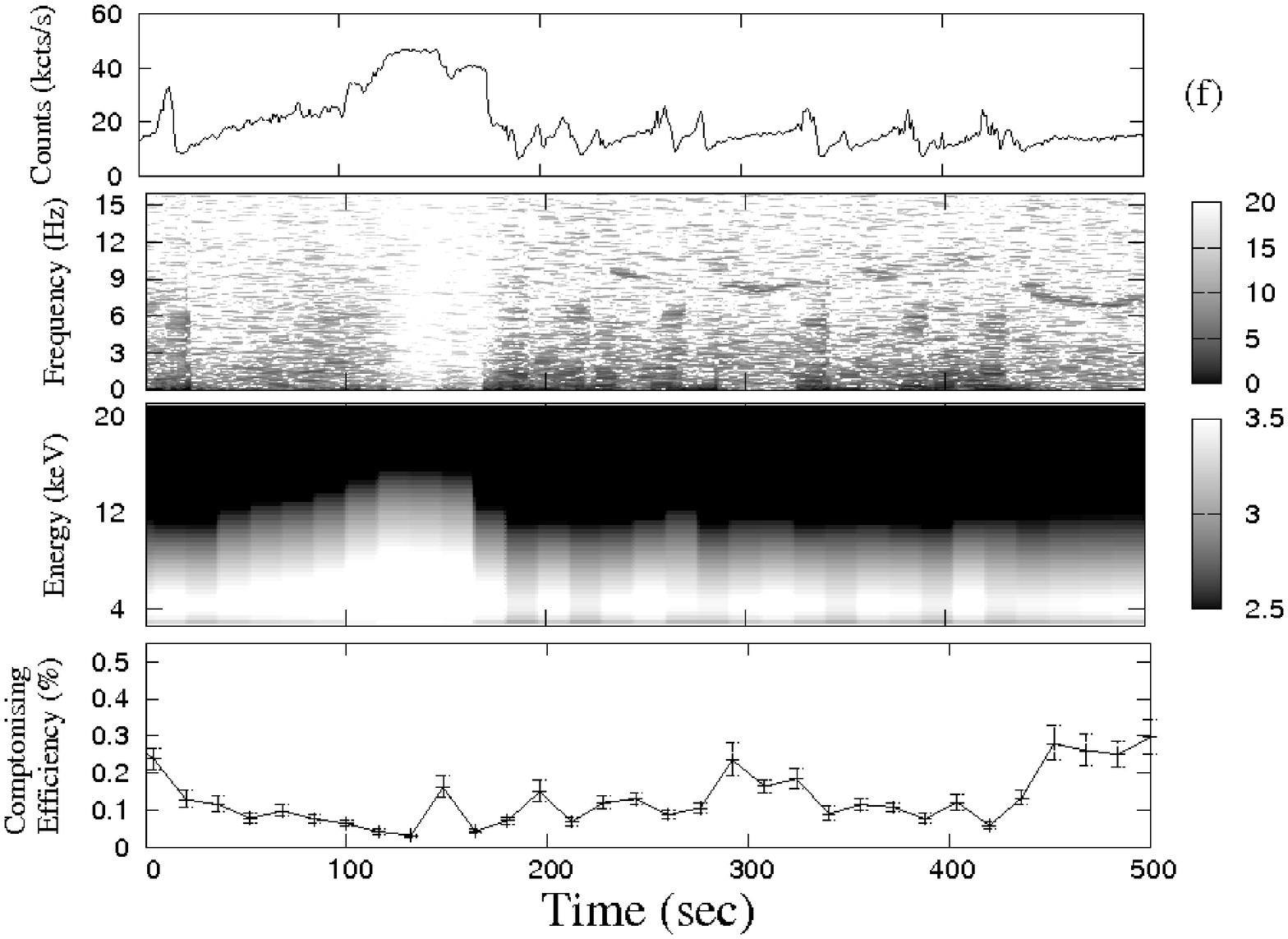}\\
}
   \centering
{
   \includegraphics[angle=270,width=6.7cm]{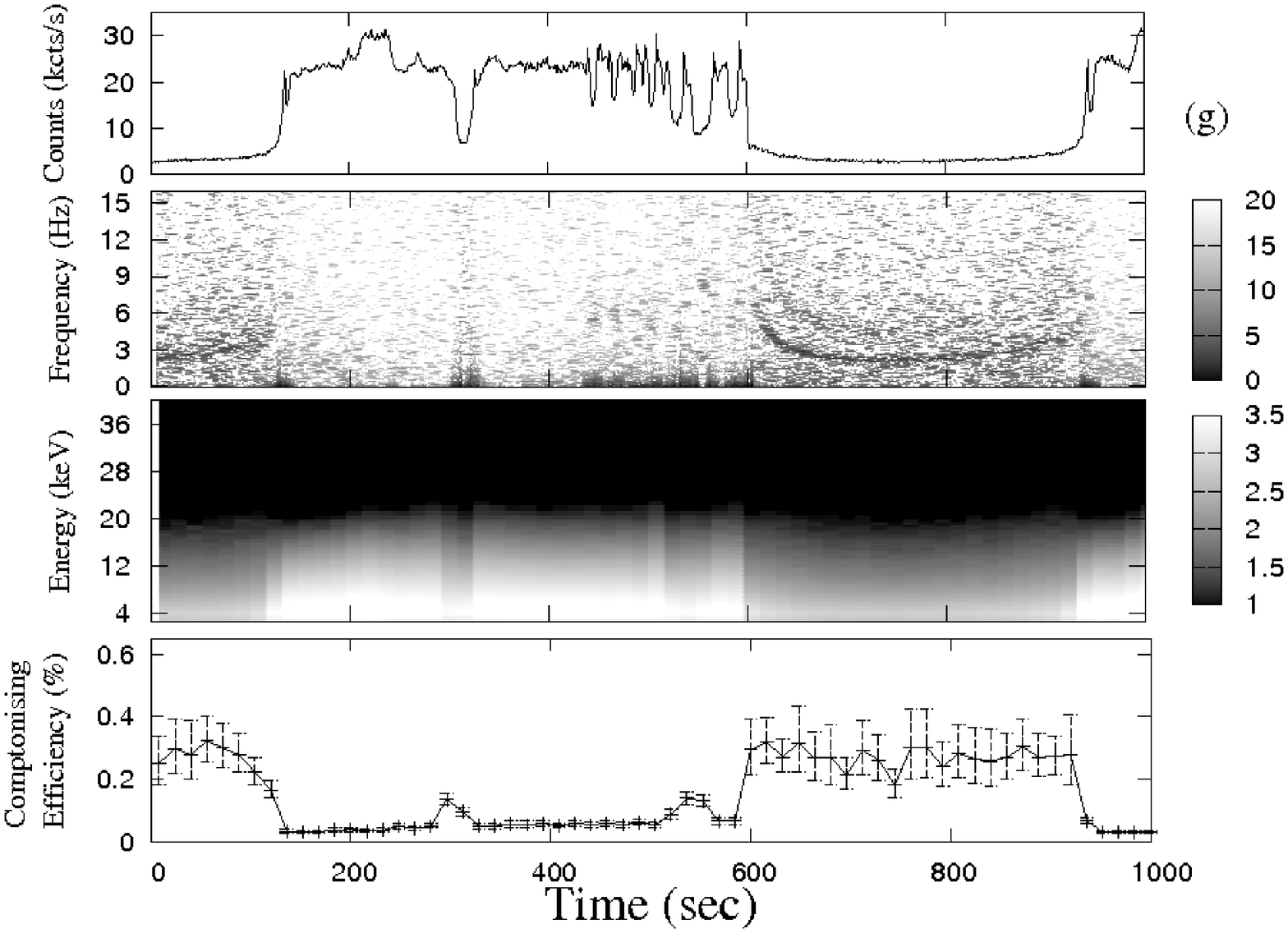}
   \includegraphics[angle=270,width=6.7cm]{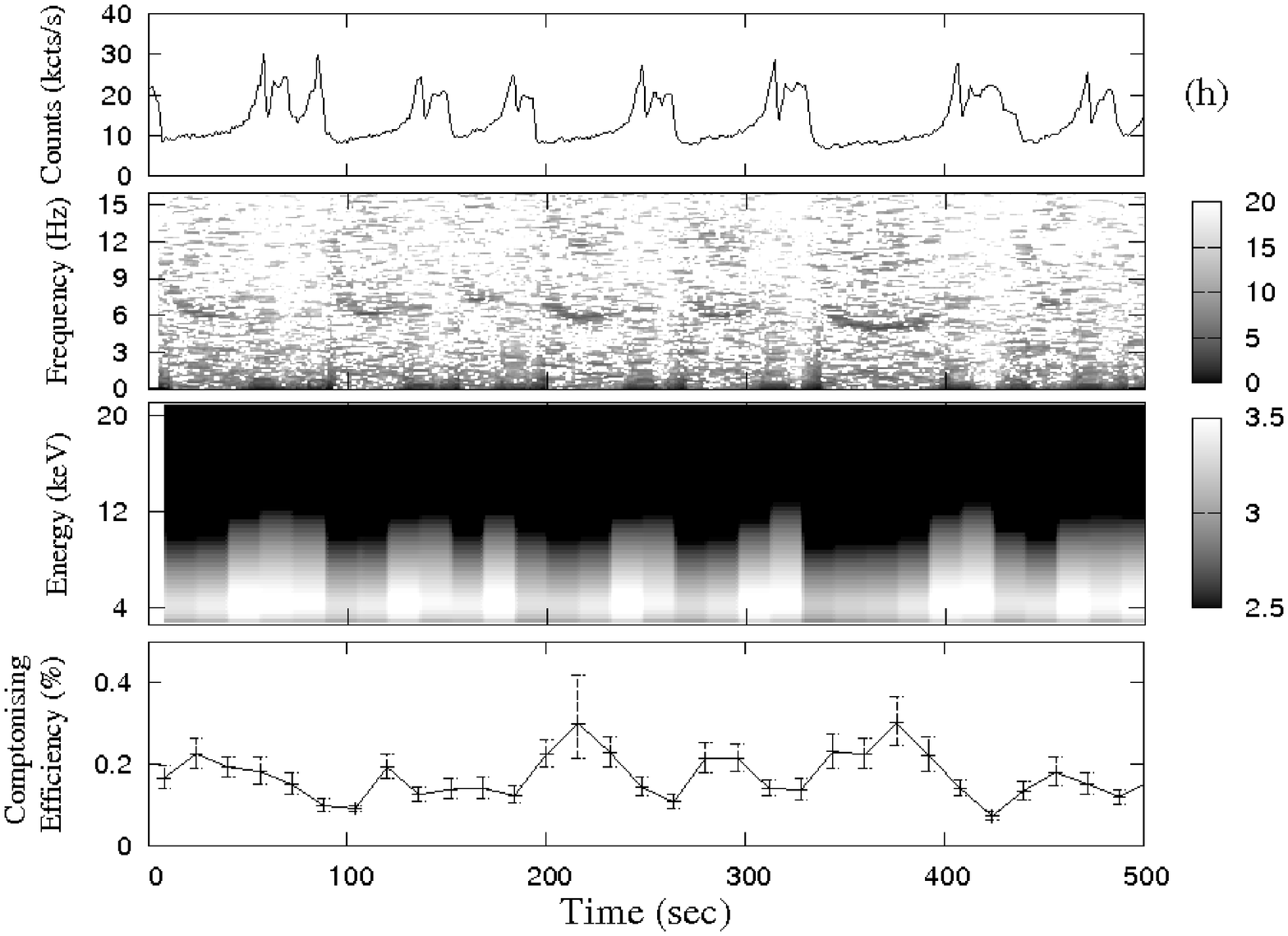}\\
}
   \centering
{
   \includegraphics[angle=270,width=6.7cm]{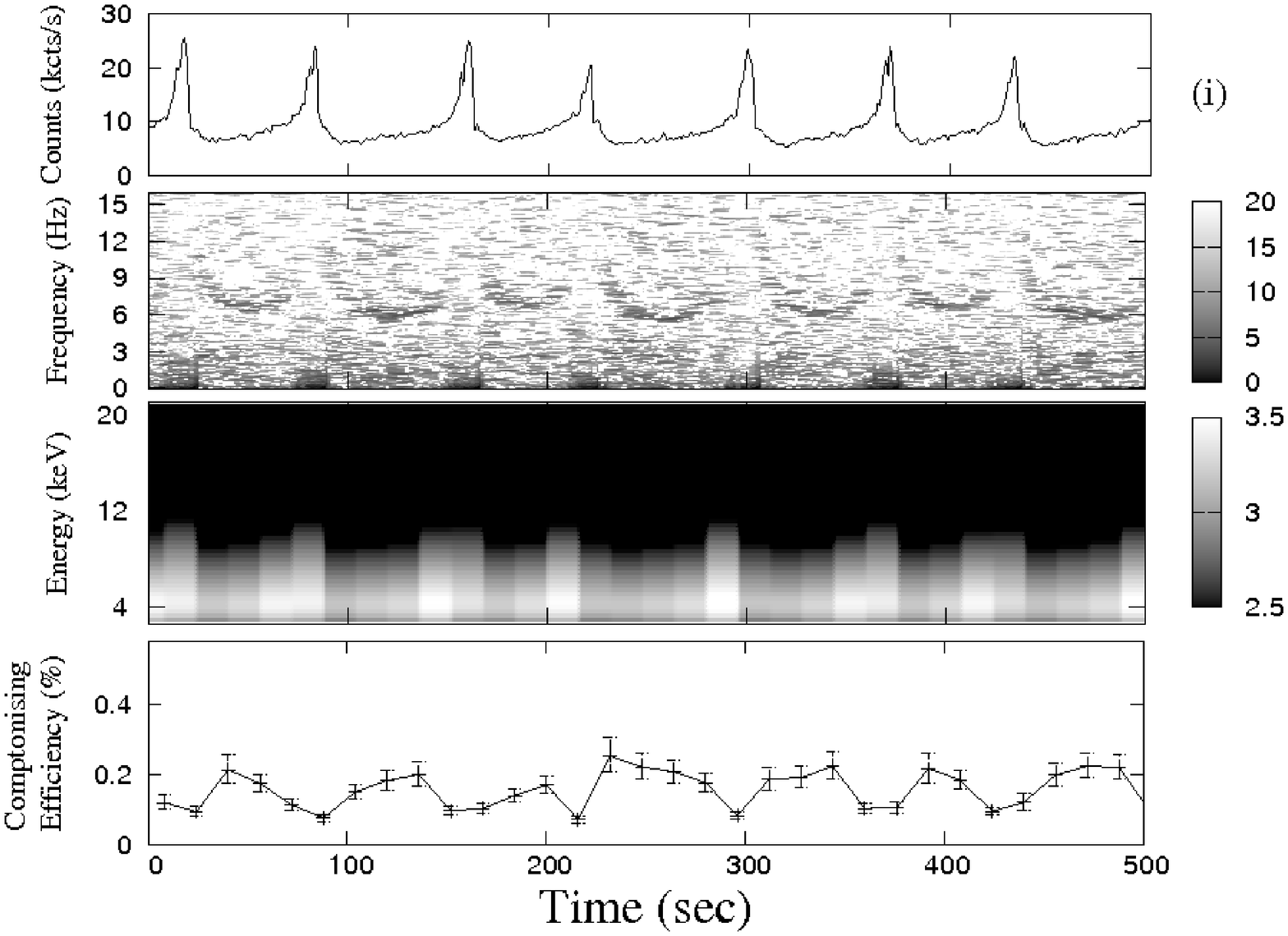}
   \includegraphics[angle=270,width=6.7cm]{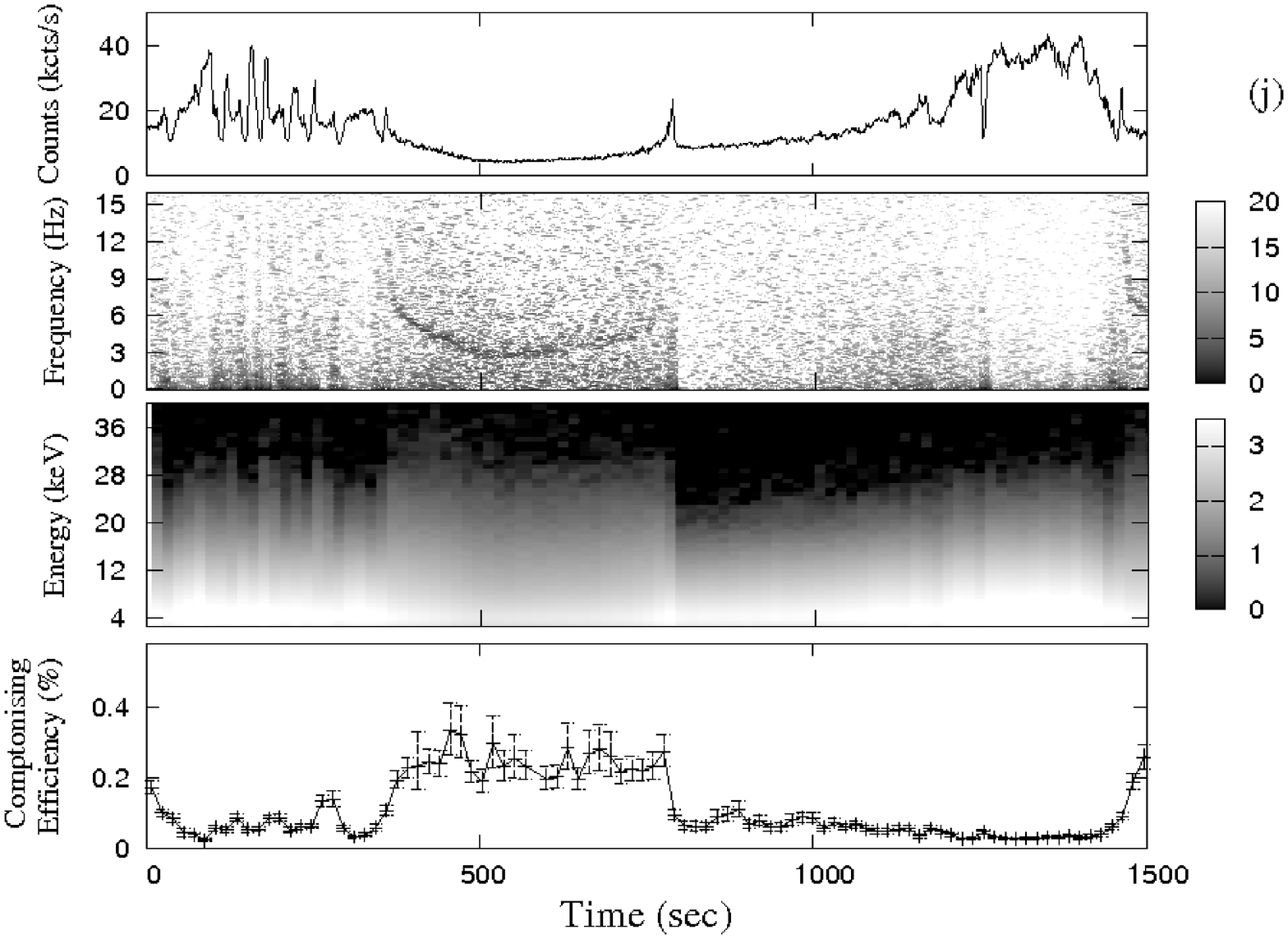}\\
}
   \label{fig3}
   \end{figure*}
%
%
   \begin{figure*}
   \centering
{
   \includegraphics[angle=270,width=6.7cm]{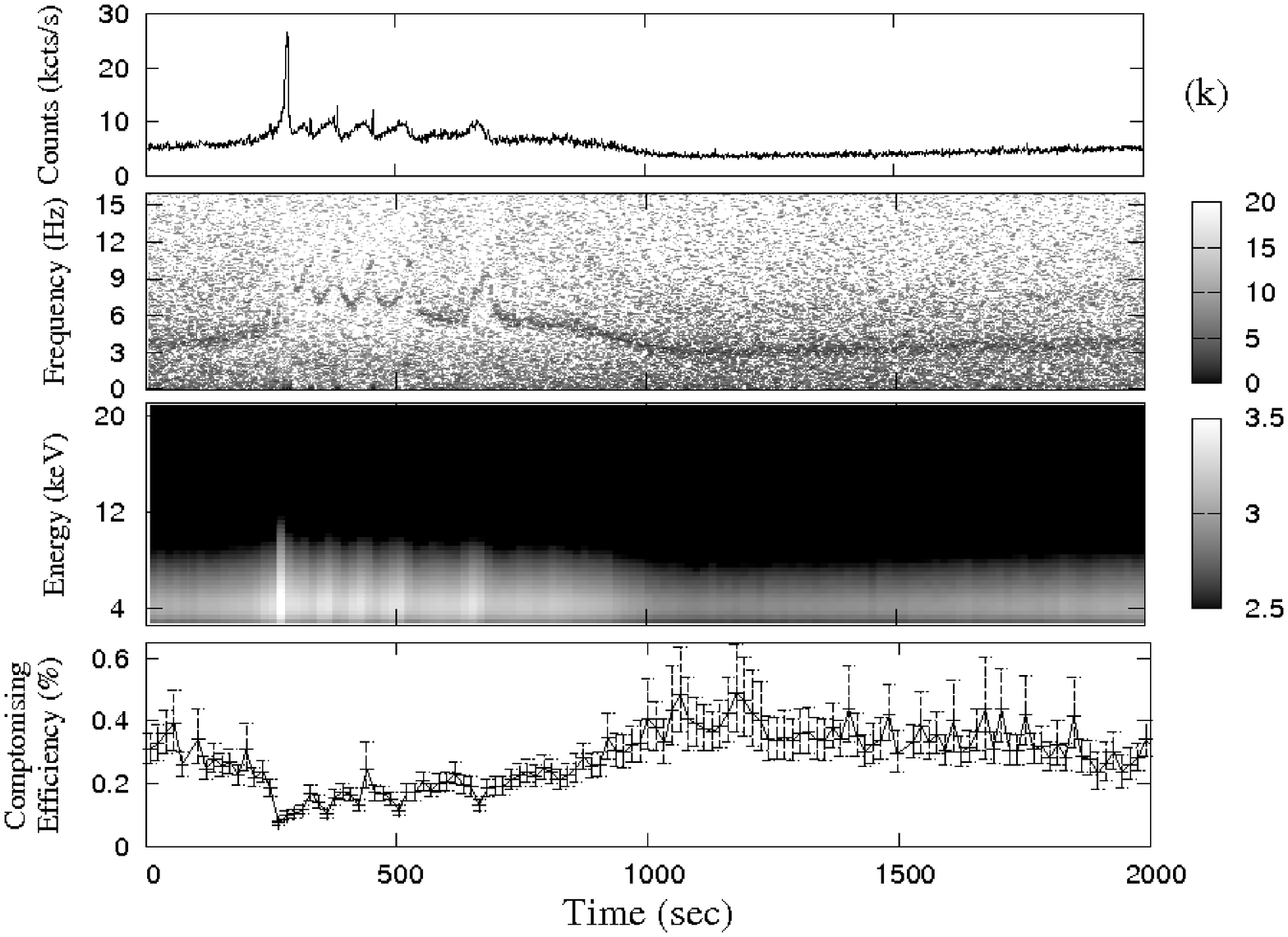}
   \includegraphics[angle=270,width=6.7cm]{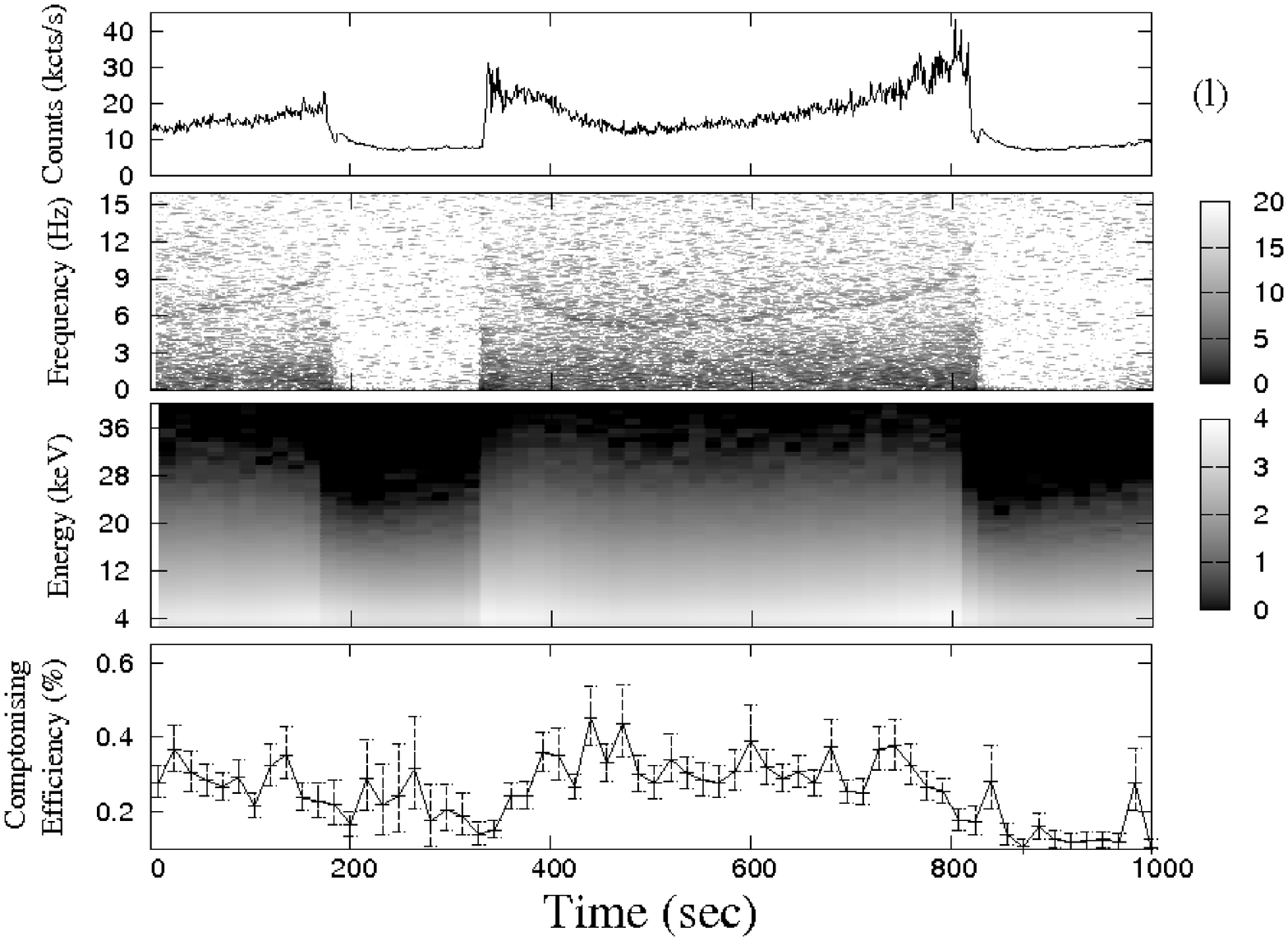}\\
}
   \centering
{
   \includegraphics[angle=270,width=6.7cm]{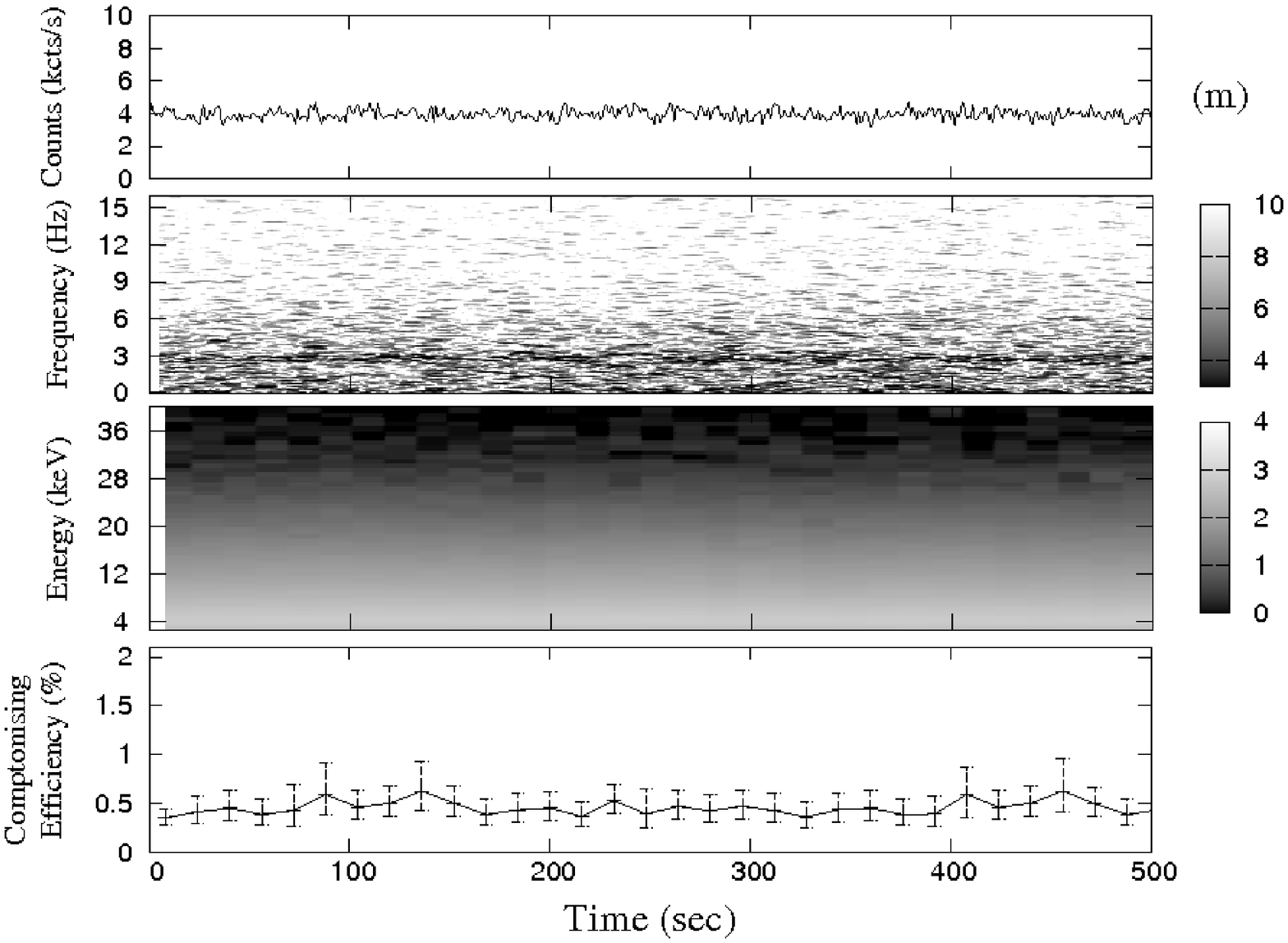}
   \includegraphics[angle=270,width=6.7cm]{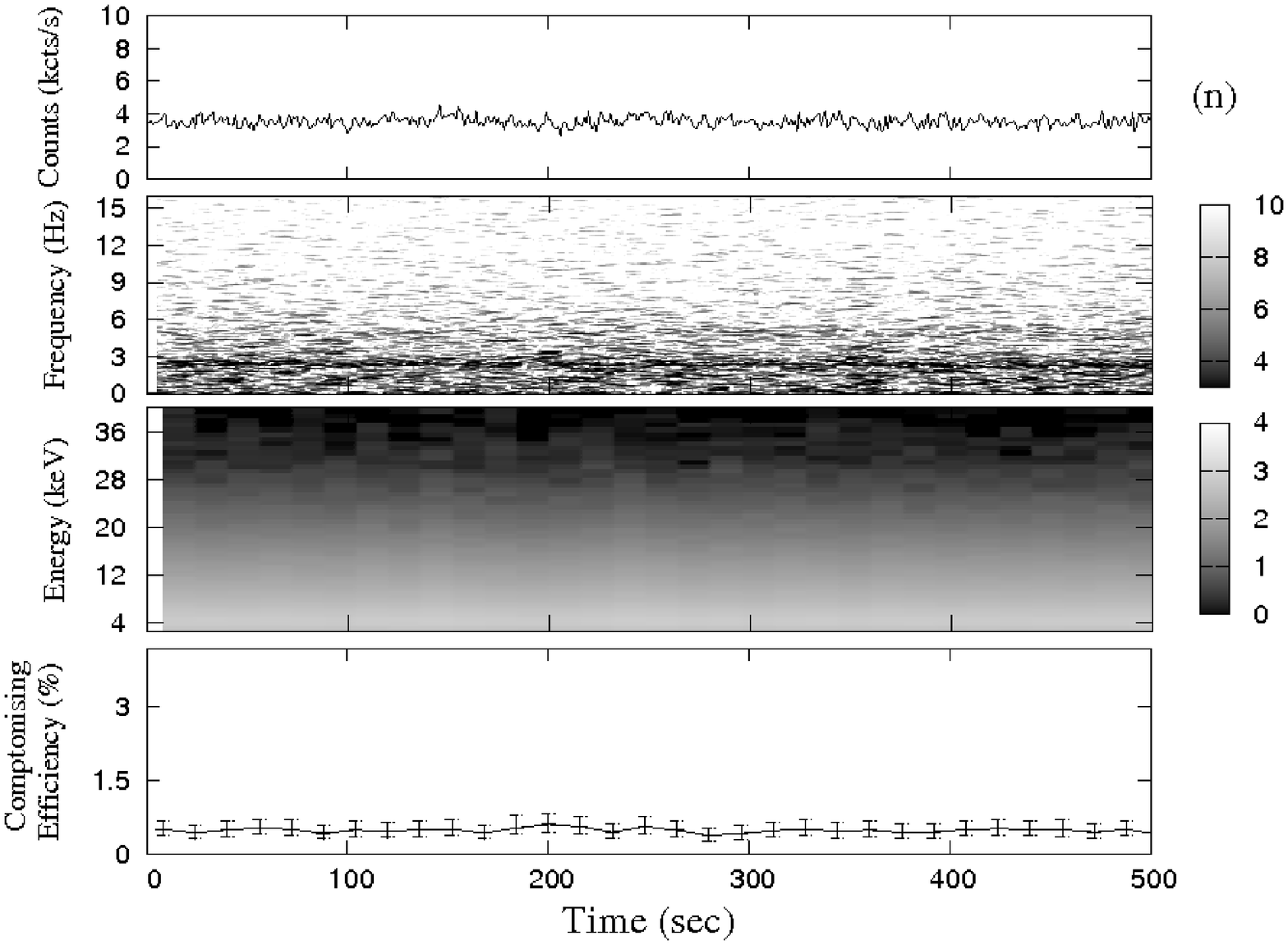}\\
}
   \centering
{
   \includegraphics[angle=270,width=6.7cm]{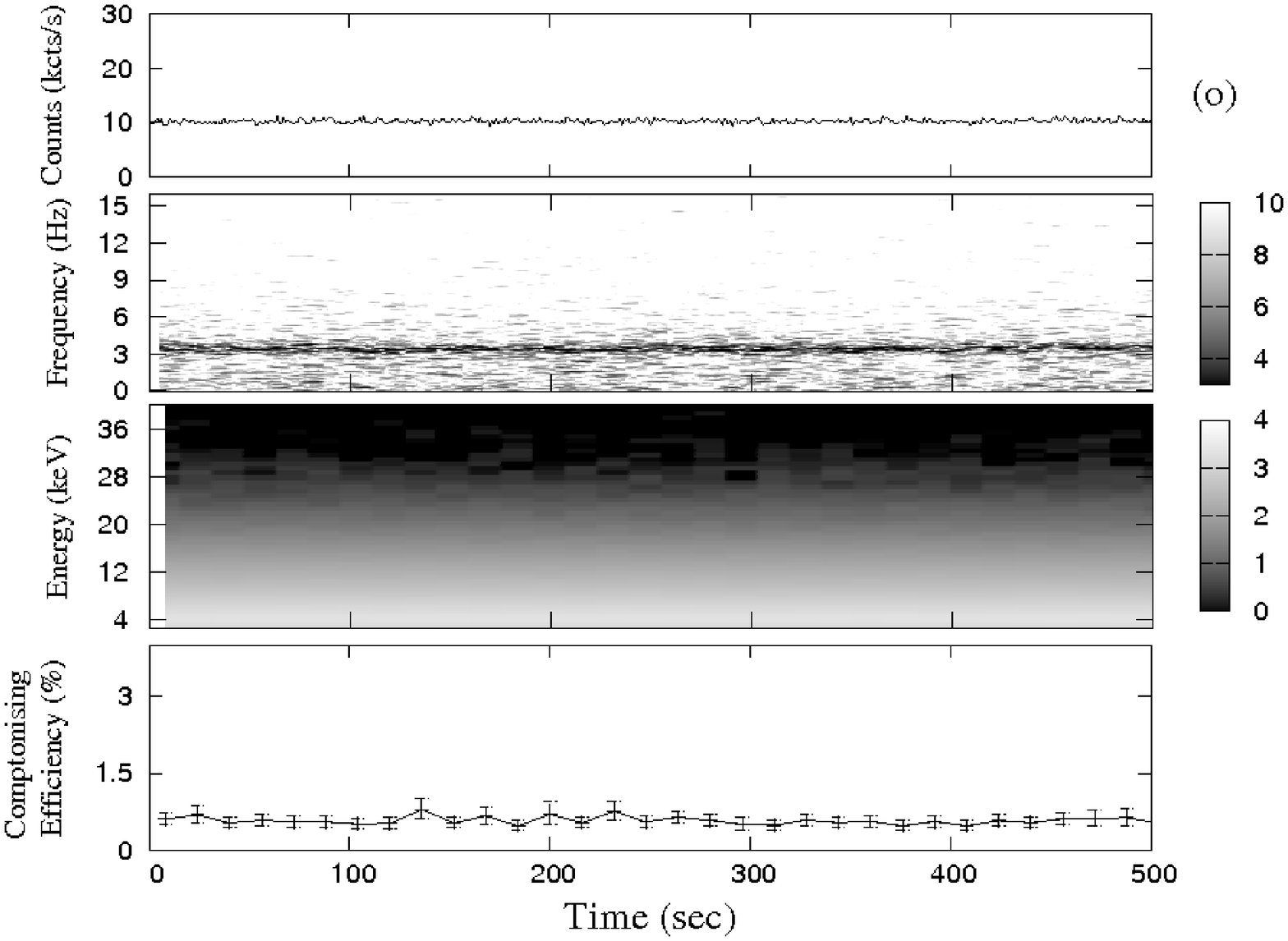}
   \includegraphics[angle=270,width=6.7cm]{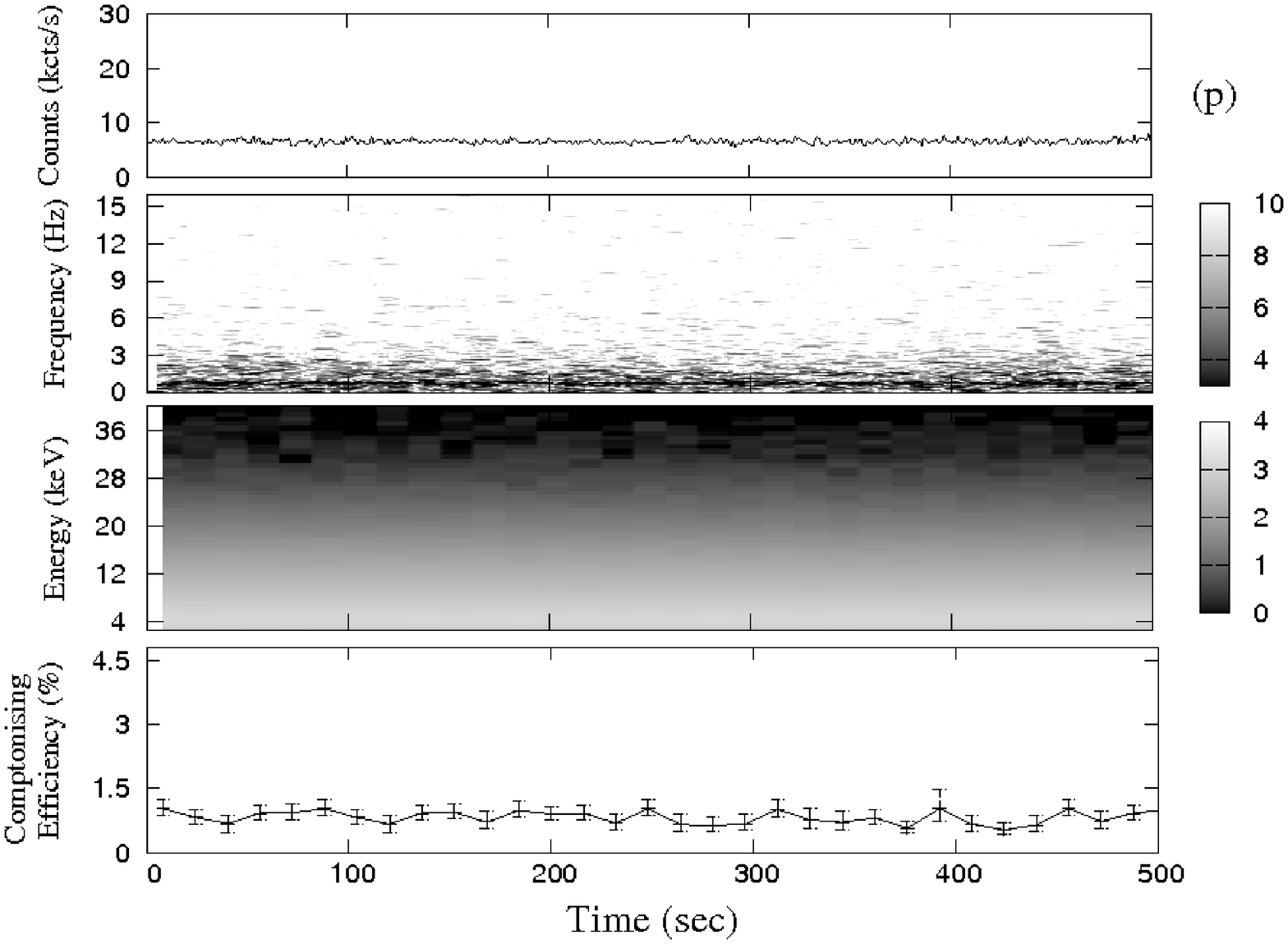}\\
}
\caption{
Results of the analysis of (a)  I, (b) II, (c) III, (d) IV,(e) V, (f) VI,
 (g) VII, (h) VIII, (i) IX, (j) X, (k) XI, (l) XII,
 (m) XIII, (n) XIV, (o) XV, (p) XVI classes are shown.
First Panel: Light curve in 2.0-40.0 keV range, Middle Panel: 
Dynamic PDS of the light curve. Clear evidence of low frequency noise and 
QPOs are seen. Third panel: Dynamic PCA spectrum showing subtle variations of the 
spectral characteristics with time. Bottom Panel: Comptonizing Efficiency (CE) in 
percent, obtained from $16$s binned data.
}
   \label{fig3}
   \end{figure*}

\subsection{Class No. I}

The Class I data shows a very short time scale variability with the presence of short scale
dips in its light curve. The results are shown in Fig. 3a.
Dynamic PDS shows no signature of QPOs and the spectrum is mostly soft in nature.
The soft photon number varies at around $330.24$ kcounts/sec while the Comptonized photon number
varies around $0.16$ kcounts/sec. The CE is only around $0.05$\%. Thus in this class
very few black body photons are intercepted by the CENBOL and correspond to a situation
similar to that in Fig. 2b. 

\subsection{Class No. II}

In Fig. 3b, we show the result of our analysis of the data that belong to the Class II. In this case,
there is an absence of QPO in the dynamic PDS and the spectrum is
soft, mostly dominated by the blackbody photons. The CE is around $0.06$\%, indicating a
situation similar to Class I.

\subsection{Class III}

The class III data appears to be less variable with a distinct and repeated 'dip' like features at a 
time gap of a few seconds. The result is shown in Fig. 3c. The class is a steady soft state 
with no QPO visible in the PDS. The CE varies around $0.06$\% indicating the low 
interception of the soft photons with the CENBOL.

\subsection{Class IV}
The results of the analysis of the data of $1000$s in class IV
is shown in Fig. 3d. The photon count seems to be steady at around $10000$ 
counts/sec for most of the time but sometimes for a duration of 
around $100$ sec the photon count is decreased to about $2000$ counts/sec. 
In the whole class, no QPO is observed and whenever the photon count is 
low, the spectrum appears to become harder. The CE remains low at around $0.07$. 

The four classes I-IV belong to the softer class. The CE is very low, even when the
Keplerian rate is high (spectrum is dominated by soft photons). This means that the CENBOL is 
very small in size and hence the QPOs are also absent. 
In our picture, these cases would correspond to
that in Fig. 2b, only accretion rates vary. From the count rates, it appears that the disk rates are 
intermediate in classes I and II while it is high in class III. The sub-Keplerian rate is low in 
class II, but is intermediate in classes  I and III.

\subsection{Class V}
The result of our analysis for $500$s of data of class V is shown in Fig. 3e. 
In this class, the spectrum, while remaining soft shows a considerable 
fluctuation. CE remains low at around $0.05\%$ to $0.15\%$. It goes up when the spectrum is 
harder. This shows that the Keplerian rate may be changing rapidly and the
CENBOL is not really formed, i.e., the sub-Keplerian flow has very low energy and/or
angular momentum (C90).  

\subsection{Class VI}

The analysis of $500$s of class VI data is shown in Fig. 3f. 
The fluctuations of the photon count rate and the spectrum are erratic.
QPOs are visible and CE is higher only when the spectrum is harder. CE varies 
between $0.04$\% to $\sim 0.3$\%. 

\subsection{Class VII}

The result of our analysis of $1000$ seconds of data of class VII is shown in Fig. 3g. 
For the first $200-600$ sec the photon number varies between $10000$ and $30000$ counts/sec.
The black body photon rate of the simulated spectrum 
is around $300$ kcounts/sec and Comptonized photon rate is around $0.19$ kcounts/sec.
This class is a mixture of the burst-off and burst-on states. When the photon count is higher, 
the spectrum is softer and the object is in the burst-on state
and when the photon count rate is lower, the spectrum is harder and object is in burst-off state.
The CE factor varies between $0.05$\% to $\sim 0.1$\%. Here too, the physics of varying the 
Comptonizing region (and thus the CE factor) is similar to what is seen in classes VIII-IX below.
At the end of the $400$s span when the burst-off 
state starts, a strong spike in both the CE factor and QPO frequency are observed,
indicating a sudden flaring in the CENBOL configuration.

 
\subsection{Class  VIII}

An analysis of a $500$s chunk of the data of class VIII of observation 
is shown in Fig. 3h. In the class VIII, the photon counts 
become high $\sim 30000$/sec and low ($\sim 10000$/sec) aperiodically at an interval of about $50-75$s.
In the low count regions, the spectrum is harder and the object is in
the burst-off state. Distinct QPOs are present and at the same time, the Comptonizing efficiency 
is intermediate, being neither as high as in the class no. XIII-XVI, nor as low as in class I-III. 
In CENBOL picture, low frequency OPOs are indicators of the shock oscillations (CM00).
As discussed before, in this case the 
strength of the shock is intermediate and produces strong outflows (CM00). 
In the fitted spectrum, the Keplerian photon varies between $150$ to $450$ kcounts/sec 
and Comptonized photon varies between $0.39$ to $0.21$ kcounts/sec.
The CE factor becomes high at $0.25$\%. The CENBOL is cooled down. After the matter which fell back on the
CENBOL is totally drained, the burst-off state is resurfaced with a lower CE (less than $0.1\%$). 

\subsection{IX class}

The result of our analysis of the class IX data is shown in Fig. 3i. This class contains a 
cyclic variation of the hard and soft photons with roughly $75$s of periodicity. 
The blackbody photon count in the 
simulated spectrum varies between $350$ to $420$ kcounts/s and Comptonized photon 
counts vary around $0.95$ to $0.34$ kcounts/s. In the harder state (low count regions), the CENBOL is prominent 
and the Keplerian component is farther away. Thus the QPO is prominent there also.  A gradual softening 
of the hard spectrum indicates that the Keplerian disk is moving towards the black hole
and the CENBOL is becoming smaller in size while still retaining its identity. 
Keplerian photon interaction increases with CENBOL size and CE rises to a maximum of $0.25$\%. 
At the peak of the light curve, the CENBOL, which is also the base of the jet is cooled down
due to the increased optical depth. 

\subsection{Class X}

The result of analysis of $1600$s of Class X data is shown in Fig. 3j. In the first phase of $600$ sec, the 
quasi periodic variation of photon counts takes place in gradually decreasing period from $10000$
counts/sec to $30000$ counts/sec. In this phase, the spectrum is soft and the QPO is seen only when 
the photon count is low. Here, the CENBOL is visible when the Keplerian flow is going far away
from time to time. The CE varies around $0.14\%$. 

In next $800$ sec the spectrum is harder and the distinct variation of QPO frequency (from 
$12$ Hz to $3$ Hz) indicates the variation of the shock location. However, in this phase,
CE is varying around $0.27\%$. This means that outflow is taking an active
role in intercepting the soft-photons.

\subsection{Class XI}
Class XI is an intermediate class in which QPO is always observed, 
the QPO frequency is generally correlated with the count rate (as in
XIII-XVI classes discussed below).
Since this class displays a long time variation, we analyze the data of $2000$s.
Fig. 3k shows the result. 
The CE varies between $0.05$\% to $\sim 0.6$\%.
The spectrum shows that the Keplerian rate is not changing much, but the sub-Keplerian flow
fluctuates, perhaps due to failed attempt to produce sporadic jets. The CE varies considerably
and so does the CENBOL.

\subsection{Class  XII}

The result of the analysis of a $1000$s data of class XII is shown in Fig. 3l. This
class can be divided in two regions depending on the photon count rates. In the soft dip region the 
photon count rate is lower than $10000$ counts/sec. In this region, the spectrum is softer
and the CE amount of interaction is around $0.17\%$.

In the other (hard dip) region, say between $350$s and $820$s,
the photon count is higher and varies from $10000$ counts/sec to 
$30000$ counts/sec. In this region, CE reaches a high value of $0.28$\%. 
The QPO is also present. Here, the CE factor,
which is linked to the geometry of the Comptonizing region, is anti-correlated  
with the QPO frequency (and correlated with shock location, i.e., CENBOL size). 
In the third panel, where the dynamic spectrum is drawn, we note that between the soft dip and
the hard dip, the low energy photon intensity is not changing much, while the intensity of 
spectrum at higher energies is higher. This indicates that the sub-Keplerian rate is high
in the hard dip state.

\subsection{Classes No. XIII \& XIV}

XIII and XIV classes have low soft X-ray fluxes with less intense radio emissions. 
Dynamic PDS shows a distinct QPO at around $3$ Hz in class XIII and around $5$ Hz in class XIV. 
The X-ray photon is significant up to $40$ keV with a flatter power-law slope, which signifies that the 
source belongs to a hard state. The results are given in Fig. 3m and Fig. 3n.
The Comptonizing efficiency in XIII and XIV classes is around $0.4-0.5$\%.


\subsection{Classes no. XV \& XVI}

The classes XV and XVI correspond to the radio-loud
states, whereas  classes XIII and XIV are in radio-quiet states (Vadawale et. al., 2001). 
In both the cases, the dynamic PDSs show a strong QPO feature around 
$4$ Hz and $0.8$ Hz respectively throughout the particular observation and the spectrum is dominated 
by the hard power law. Results are shown in Fig. 3o and Fig. 3p. Around $0.77$\% of 
the soft photons are intercepted by the CENBOL in XIII class, whereas it is $0.88$\% in XVI class.
In our picture, the CENBOL (Fig. 2a) oscillates to produce the observed QPOs (Chakrabarti, 1997; CM00). 
Generally, XIII-XVI classes are in harder states and in the language of 
two component model (CT95, C97), the Keplerian rate is low and unable
to cool the CENBOL and jet combined which are produced by 
the sub-Keplerian flows. Whether the jets would be produced or not depends on the
shock strength (C99).

\section{Comptonizing efficiency in different classes: an unifying view}

In the above sections we have analyzed the data of all the sixteen classes. 
We presented the light curves, the dynamical energy and power density spectra
and the ratio of the power-law photons to the black body photons from the simulated fit
of the original spectra.
In the literature, there has been so far no discussion on the sequence in which the class transitions
should takes place. Also, there is no discussion on
which physical properties of the flow, the characteristics of the 
variability classes depend. In the present paper,
we analyzed many aspects of the variability classes 
which may be understood physically from the two component advective disk paradigm.
Since CE behaves differently in various classes, it may play
an important role in distinguishing various classes. This CE
is directly related to the size and optical depth of the Compton cloud 
which determines the fraction of the injected soft photon which are intercepted 
by the Comptonizing region.
   \begin{figure*}
   \centering
   \includegraphics[angle=-90,width=7cm]{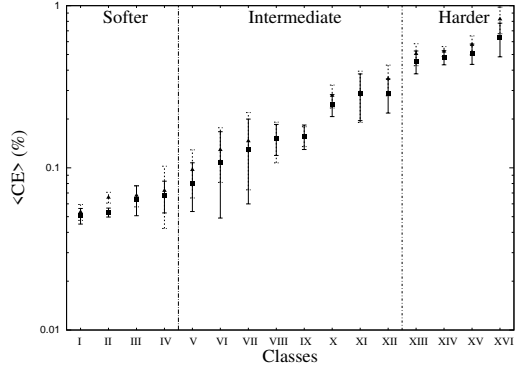}
\caption{Variation of average Comptonizing efficiency ($<CE>$) for different variability classes 
of GRS 1915+105. The error-bars were calculated from the excursion of CE in a given class. Filled squares and 
triangles represent $<CE>$ for two sets of variability classes. Generally softer classes have smaller $<CE>$ and
harder classes have higher $<CE>$.
\label{seq}}
     \end{figure*}

Though CE varies very much in a given spectral class, it is instructive to compute the average  
CE in a given class (averaged over a period characterizing the class). We denote this as $<CE>$. 
In Fig. 4, we present the variation of log($<CE>$) in various classes. We placed error bars
also in all the average values which are at 90\% confidence level.
We arranged  arbitrarily defined sixteen classes in a manner so that $<CE>$s are monotonically increasing.
This gives rise to a sequence of the variability class shown in the X-axis. In classes X and XII
we did not average CE over both hard and soft
regions, since it is believed that the soft regions are produced due to sudden
disappearance of the hard region (e.g., Nandi et al. 2001). 
Thus, while placing them in the plot, we considered the average over the 
burst-off (harder) state only. To show that the sequence drawn is unique, we plotted the  
average values of two sets of data of all the classes. The average of 
the first set is drawn with a dark square sign and the average of the 
second set is drawn with a dark triangle sign. The individual error bars are 
drawn with solid and dashed lines respectively. In both the sets, the sequence is identical
with softer classes to the left and harder classes to the right. Based on the nature of the 
variations of CE, we divided these classes into three groups corresponding to three
types of accretion shown in Fig. 2(a-c). Classes I to IV belong to `softer' state
and classes XIII-XVI belong to `harder' state. The rest of the classes belong to the `intermediate' state.

The above findings of alignment in average CE variation shows that it plays a significant role
in distinguishing the variability classes. Moreover, as we discussed already, the variation 
in CE for a given class is due to the softening and hardening of the spectra in a much shorter 
time-scale and thus is related to `local' physical processes, such as interaction with winds 
and outflows, which, in turn, depends on the existence or non-existence of CENBOL, i.e., 
the spectral states. 

   \begin{figure*}
   \centering
   \includegraphics[angle=-90,width=7cm]{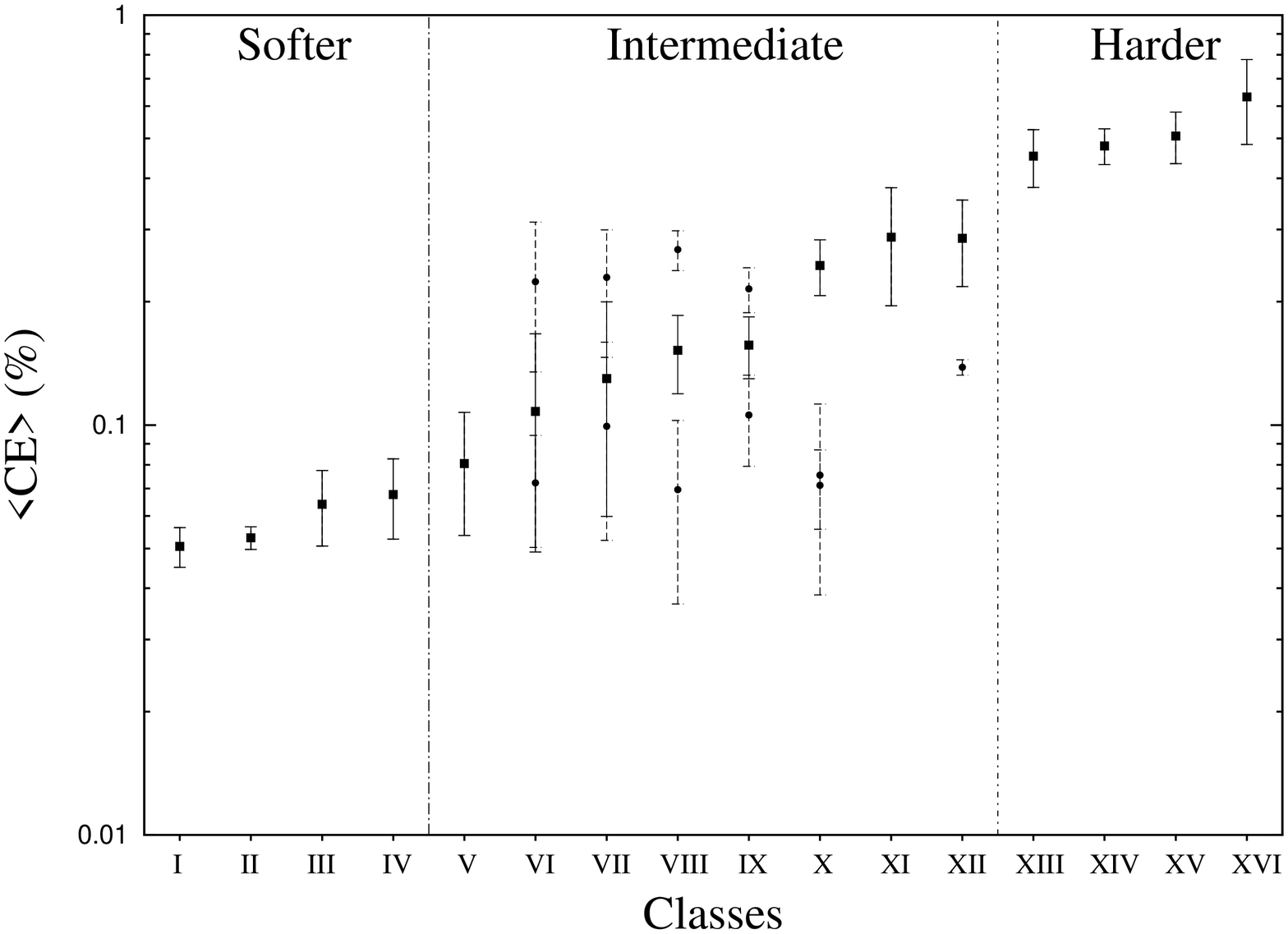}
   \includegraphics[angle=-90,width=7cm]{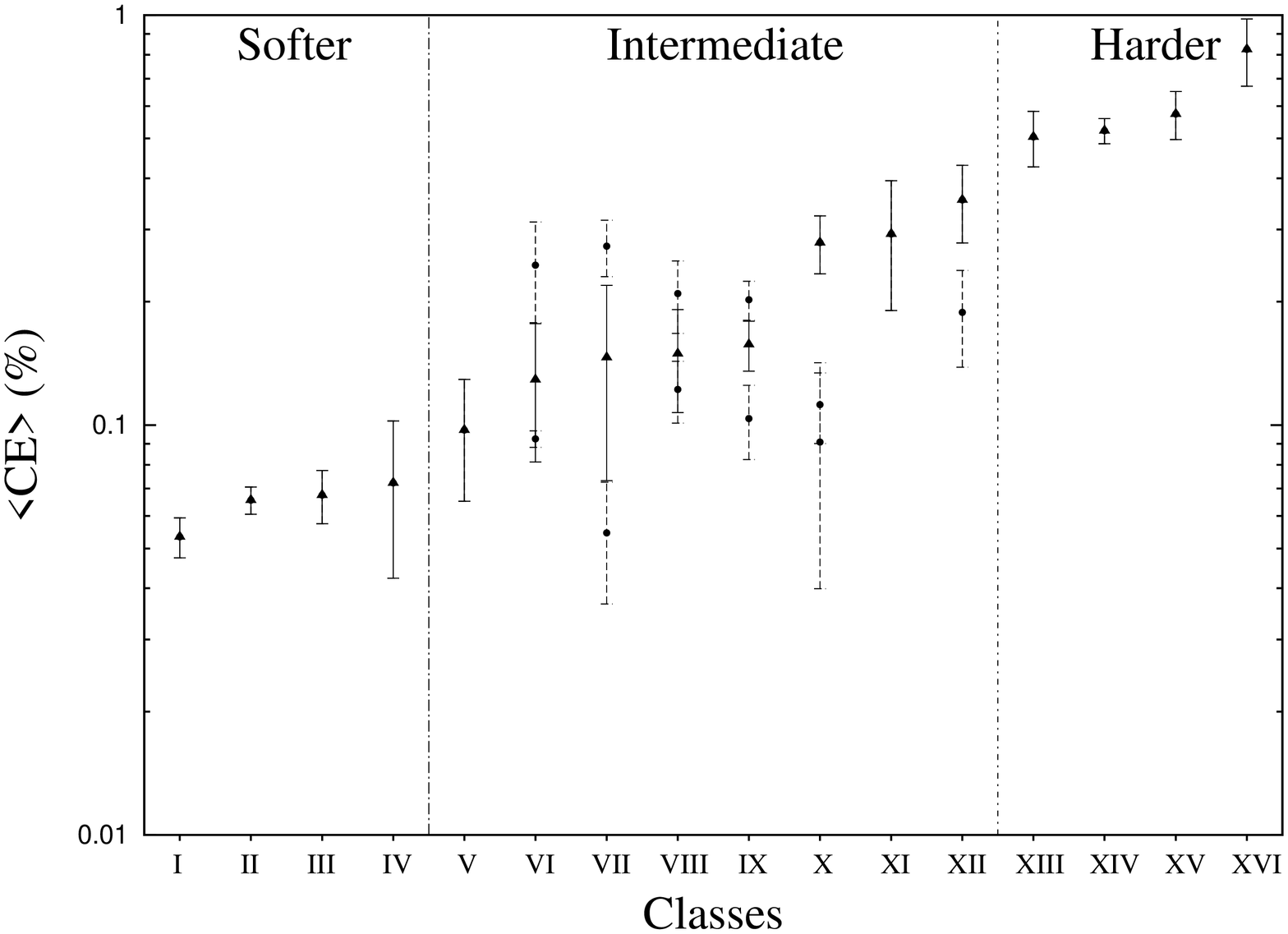}
\caption{Variation of averaged Comptonizing efficiency ($<CE>$) for different variability 
classes of GRS 1915+105.
In the left panel, we show the results for the data set plotted with filled squares in Fig. 4,
and the right panel, we show the results of the
data plotted with filled triangles in Fig. 4. The averages in hard
and soft chunks are placed separately with filled circles. In class X
averages of three chunks (burst on, pre-spike and post-spike) are plotted with filled
circle, filled box and filled circle respectively.
The sequence is found to remain the same in both sets of data. We placed classes X and XII
according to $<CE>$ in the hard (pre-spike) region since the softer regions
are believed to be produced due to totally different physical processes.}
\label{seq2}
     \end{figure*}

In Fig. 5 (a-b), we plotted the results of these two sets of analysis separately. 
We show separate averages over CE in the burst-off and burst-on states 
(as two dark circles) when they are present as well as the global average 
over CE in a given class by filled squares (except classes X and XII where the physics 
is different and it is meaningless to talk about the overall average; and hence only 
average over the harder region is plotted.).

In Chakrabarti et al. (2004) and Chakrabarti et al. (2005), it was mentioned 
that although many observations were made of GRS 1915+105, only a few cases, 
direct transitions were observed. In particular, using Indian X-ray Astronomy 
Experiment (IXAE) they showed direct evidences of $\kappa \rightarrow \rho$ (VIII $\rightarrow$ IX), 
$\chi \rightarrow \rho$ (XIII $\rightarrow$ IX), $\chi \rightarrow \theta$ (XIII$\rightarrow$ XII) and $\rho \rightarrow \alpha$ (IX$\rightarrow$ XI) transitions in a matter of hours. In Naik et al. (2002), IXAE
data was used to argue that $\rho$ class variability could have changed 
to $\chi$ class via $\alpha$ class (i.e., IX $\rightarrow$ XI $\rightarrow$ XII). In Nandi et al. (2001), it was 
shown that the $\theta$ class (Class XII) is rare and the observed soft-dip is perhaps due to the disappearance 
of the inner region by magnetic rubber band effect. Accepting that 
class XII as anomalous, we find from Fig. 4 that the observed 
transitions reported in Chakrabarti et al. (2004, 2005) and Naik et al, 2002 are `naturally' explained. 
For instance, $\chi \rightarrow \rho$ ($\theta$ and $\beta$ being anomalous and intermediate
$\alpha$ has been reported by Naik et al. 2002), $\rho \rightarrow \alpha$
($\beta$ being anomalous) and $\kappa \rightarrow \rho$, 
are expected from our analysis. Similarly, we can claim that there should not be any transitions
such as I $\rightarrow$ IX; III$\rightarrow$ VII; for example. We believe that if we
carry out spectro-photometry of GRS 1915+105 continuously (Chakrabarti et al. 2008), then
we may be able to catch the transition from one type to another more often and verify if the 
sequences we mentioned here need further refining. 

\section{Conclusions}

In this paper, we have analyzed all known types of light curves of the enigmatic black hole GRS1915+105
and computed the dynamical nature of the energy and the power density spectra. We did not characterize
these classes by conventional means, such as using hardness ratios defined in certain energy range since
such a characterization does not improve our view about the physical picture. Furthermore, characterization
using certain energy range is possible only in a case by case basis, and is not valid for 
the black holes of all masses. Instead, we asked ourselves  whether we can distinguish one class from another 
from physical point of view purely in a model independent way. 
We observed that independent of what the nature of the Compton cloud is,
the weighted mean $<CE>$ of Comptonizing efficiency (CE) obtained 
every 16 seconds of the binned data, increases monotonically as the class varies. 
This pattern we find was verified with two sets of data covering all the variability classes. 
So, <CE> is not arbitrary -- it is characteristics of a class. Based on the values of <CE>, it is observed that 
the classes belong to three states: Classes I-IV in softer states, Classes V-XII in intermediate
states, and Classes XIII-XVI belong to the harder state.

When the weighted average value of CE is monotonically arranged, we obtain a sequence which appears to be
followed during the actual transitions. Indeed, when comparing with available data of PCU of RXTE and 
the Indian payload IXAE onboard IRS-P3, the transitions from one class to another as reported in the 
literature do follow our sequence. Given that a large variation of CE occurs in a given variability
class, it is puzzling why the sequence of the classes obtained by us should follow changing the
CE values {\it averaged} over the whole class. It is possible that a specific value of $<CE>$
actually forces the system to be in a given class, just as a parameter such as wind speed
decides the mean angle of an oscillating pendulum. The excursion of CE in that class could be 
due to totally different physical process and not necessarily due to interception of soft photons
by the CENBOL alone. May be the mass-loss rate of CENBOL is playing a role, which in turn depends on the
shock strength.

In the two component model of the Chakrabarti-Titarchuk (CT95), the CENBOL and associated outflow 
play the role of the Compton cloud. In this model, the increase in Compton efficiency 
can be affected in several ways: (i) by increasing the shock location increases the size of the 
CENBOL which intercepts larger number of soft photons
and/or (ii) by increasing the accretion rate of the sub-Keplerian component, which increases
the optical depth and scatter more soft photons to produce power-law photons. 
In CT95 and C97 it was shown that harder states are produced by 
both the effects mentioned above. Indeed, we find that the classes with 
harder states have more CE. Shifting of the shock locations is possible by changes in viscosity
or changes in cooling rate in the post-shock region. The
time scales of such effects in a sub-Keplerian flow could take hours.  Time scale of 
changing the global sub-Keplerian flow rates could be comparable to the free-fall time from the outer
edge, i.e., of the order of a day or so. On the other hand, in Classes VII and VIII etc.,
the CE changes in a matter of minutes. Such a short time variability of Comptonizing efficiency 
is possible if the base of the outflow is abruptly cooled and returned back to the disk
increasing the accretion rate of the Keplerian/sub-Keplerian rates locally (CM00).

In this paper, we find that the $<CE>$ is really important in deciding the sequence.
However, for a given class, the degree of excursion of CE in a 
given variability type must depend on another parameter, such as the outflow rate 
which in turn depends on the shock strength. A simple 
estimate (C99) suggests that the shock strength decides the outflow rate and hence the time 
taken by the base of the outflow to reach unit optical depth $\tau$ for Compton scattering (CM00). 
For a very strong shock, this outflow rate is very weak, as is evidenced by weak (few tens
of miliJansky) radio flux even for `radio-loud' classes. For intermediate shock strength 
the outflow rate is higher, and it is easier to have $\tau=1$ in a short time scale (CM00)
and fractional change in CE also becomes high.  In classes XII-XVI, we not only see
CE to be very high, the fractional change in CE is very small as well.
The aspect of classification in terms of outflow rate is being looked into. 
The analysis is in progress and will be reported elsewhere.

\begin{acknowledgements}
The work of P. S. Pal is supported by a CSIR Fellowship.
\end{acknowledgements}

\end{document}